\documentclass[twocolumn,showpacs,preprintnumbers,amsmath,amssymb]{revtex4}
\usepackage{epsfig}
\usepackage{dcolumn}% Align table columns on decimal point
\usepackage{bm}% bold math
\usepackage{pstcol,pst-grad}% PsTricks

\begin{document}

\preprint{APS/123-QED}
\title{Spanning  avalanches in  the three-dimensional  Gaussian Random
Field  Ising  Model with  metastable  dynamics:  field dependence  and
geometrical properties}
\author{Francisco J. P\'erez-Reche}   
\email{jperez@ecm.ub.es}   
\author{Eduard Vives}  
\email{eduard@ecm.ub.es} 
\affiliation{ Dept.   d'Estructura i  Constituents  de la  Mat\`eria,
Universitat  de Barcelona  \\ Diagonal  647, Facultat  de F\'{\i}sica,
08028 Barcelona, Catalonia}
\date{\today}

\begin{abstract}
  Spanning  avalanches in  the 3D  Gaussian Random  Field  Ising Model
  (3D-GRFIM)  with metastable  dynamics at  $T=0$ have  been  studied. 
  Statistical analysis of the  field values for which avalanches occur
  has  enabled a  Finite-Size  Scaling (FSS)  study  of the  avalanche
  density  to be  performed.  Furthermore, direct  measurement of  the
  geometrical properties  of the  avalanches has confirmed  an earlier
  hypothesis  that  several  kinds  of spanning  avalanches  with  two
  different  fractal dimensions  coexist  at the  critical point.   We
  finally compare  the phase diagram  of the 3D-GRFIM  with metastable
  dynamics with the same model in equilibrium at $T=0$.
\end{abstract}


\maketitle

\section{Introduction}
\label{Intro}

Avalanche  behavior \cite{Sethna2001}  has  been found  in many  first
order  phase  transitions  at  low temperatures.   When  the  external
parameters are slowly driven,  the transition, instead of occurring as
a sharp  change of  the system  properties at a  certain point  of the
phase diagram, splits  into a series of discontinuous  jumps that link
metastable  states.   This  gives  rise  to  hysteresis  loops,  whose
branches consist  of a  sequence of random  steps.  In many  cases the
sizes  of such  steps range  from microscopic  to  macroscopic scales,
distributed according  to a power-law.  The phenomenon  has been found
to       be      associated       with       magnetic      transitions
\cite{Babcock1990,Cote1991,Puppin2000},     capillary     condensation
\cite{Lilly1993,Kierlik2001},        martensitic       transformations
\cite{Vives1994a},   and   others   \cite{Casanova2004}.   A   crucial
ingredient  in  order  to  observe  such avalanches  is  that  thermal
fluctuations  are very small  compared with  the energy  barriers that
separate transformed and untransformed domains.  For this reason, such
first-order   phase   transitions   have  been   called   ``athermal''
\cite{PerezReche2001}              or              ``fluctuationless''
\cite{Vives1994,Carrillo1998}.

Within  this  context  the  3D-GRFIM  with metastable  dynamics  is  a
prototype model: the complexity of what we call ``disorder'' in a real
system is simplified into a series of quenched random fields, Gaussian
distributed with  zero mean and standard deviation  $\sigma$, that act
on  every spin  of a  3D Ising  model. In  addition, one  assumes that
temperature is zero  ($T=0$) and provides the model  with a particular
metastable  dynamics   in  order  to   study  the  evolution   of  the
magnetization $m$ when  the external field $H$ is  swept.  The details
of the dynamics  were introduced by Sethna and  coworkers a decade ago
\cite{Sethna1993,  Dahmen1993}.   The  basic  assumption is  that  the
driving field  rate is  slow enough so  that system relaxation  can be
considered  instantaneous (adiabatic  driving).  Such  relaxations are
the so-called magnetization avalanches.

After the introduction of  the model several works \cite{Perkovic1995,
  Dahmen1996,   Tadic1996,  Perkovic1999,   Kuntz1999,  Carpenter2002}
described the  basic associated phenomenology: (a) the  existence of a
disorder-induced critical point at $\sigma = \sigma_c$ associated with
the change  from a continuous  to a discontinuous hysteresis  loop and
(b) the fact that within a  large region around the critical point the
distribution  of  avalanche  sizes  $D(s)$ exhibits  almost  power-law
behavior.   Nevertheless, several  questions  remain unsolved,  mostly
related  to  the  properties  of  the spanning  avalanches  which  are
responsible  for  the  observed  macroscopic  discontinuities  in  the
$H$-$m$ hysteresis loop and in the $\sigma$-$H$ phase diagram.

More  recently,  a  FSS  analysis  of the  number  of  avalanches  and
avalanche  size distribution  \cite{PerezReche2003} has  revealed that
the  scenario is  quite  complex.   The study  was  restricted to  the
statistical analysis of the full  set of avalanches recorded in a half
loop,  irrespective of  the  field values  $H$  where such  avalanches
occur.   (The  obtained  statistical  distributions are  often  called
integrated distributions). A  detailed study as a function  of $H$ has
not been done before.

Let  us   summarize  here   the  main  results   presented  in   Ref.  
\cite{PerezReche2003}, in  order to introduce  the notation. According
to  the  dependence on  the  system size  $L$  of  the average  number
$N_{\alpha}(\sigma,L)$ and size distribution $D_{\alpha}(s;\sigma,L)$,
avalanches can  be classified  into several categories,  (the subindex
$\alpha$ stands  for the different  categories) as presented  in Table
\ref{TABLE1}.  The  scaling behavior is  written as a function  of the
variable $u$ which measures the  distance to the critical value of the
disorder $\sigma_c=2.21\pm0.02$.  Its precise definition is given as a
second-order expansion:
\begin{equation}
u=\frac{\sigma-\sigma_c}{\sigma_c}+          A         \left         (
\frac{\sigma-\sigma_c}{\sigma_c} \right )^2
\label{u}
\end{equation}
with $A=-0.2$. This expression was found to be the best choice for the
collapse of the  scaling plots.  The values of  the different critical
exponents are summarized in  Table \ref{TABLE2}, together with the new
exponents that will be computed in the present work.

\begin{table*}
\begin{center}
\begin{tabular}{lcll}
\hline
\hline
avalanche type & $\alpha$ & average number & size distribution \\ 
\hline 
\hline
non-spanning & $ns$   &    $N_{ns}(\sigma,L)    \;    \;    \;    \left    [
N_{ns}=N_{nsc}+N_{ns0} \right ]  $ & $D_{ns}(s;\sigma,L) \;\;\; \left
[ N_{ns}D_{ns}=N_{nsc}D_{nsc}+N_{ns0}D_{ns0} \right ] $ \\
critical   non-spanning  & $nsc$   &  $N_{nsc}(\sigma,L)=L^{\theta_{nsc}}
\tilde{N}_{nsc}  (uL^{1/\nu})$  &  $D_{nsc}(s;\sigma,L)=L^{-\tau_{nsc}
d_f} \tilde{D}_{3c} (sL^{-d_f},uL^{1/\nu})$\\
non-critical    non-spanning    &    $ns0$   &    $N_{ns0}(\sigma,L)=L^3
\tilde{N}_{ns0}(\sigma)$ & \\
\hline
1D-spanning & 1 & $N_1(\sigma,L)=L^{\theta}  \tilde{N}_1  (uL^{1/\nu})$ &
$D_1(s;\sigma,L)=L^{-d_f}  \tilde{D}_1  (sL^{-d_f},uL^{1/\nu})$  \\  
\hline
2D-spanning  & 2 & $N_2(\sigma,L)=L^{\theta}  \tilde{N}_2  (uL^{1/\nu})$ &
$D_2(s;\sigma,L)=L^{-d_f}  \tilde{D}_2  (sL^{-d_f},uL^{1/\nu})$  \\  
\hline
3D-spanning & 3 & $N_3(\sigma,L)\; \;  \; \left [ N_3=N_{3c}+N_{3-} \right
]     $     &      $D_3(s;\sigma,L)     \;\;\;     \left     [     N_3
D_3=N_{3c}D_{3c}+N_{3-}D_{3-} \right ] $ \\
critical  3D-spanning & 3$c$ &  $N_{3c}(\sigma,L)=L^{\theta}  \tilde{N}_{3c}
(uL^{1/\nu})$     &     $D_{3c}(s;\sigma,L)=L^{-d_f}    \tilde{D}_{3c}
(sL^{-d_f},uL^{1/\nu})$    \\     
subcritical          3D-spanning  & 3-        &         $N_{3-}(\sigma,L)=
\tilde{N}_{3-}(uL^{1/\nu})$      &     $D_{3-}(s;\sigma,L)=L^{-d_{3-}}
\tilde{D}_{3-} (sL^{-d_{3-}},uL^{1/\nu})$ \\ 
\hline
\hline
\end{tabular}
\end{center}
\caption{\label{TABLE1}  Classification  of  the  different types  of
avalanches in the 3D-RFIM according to their geometrical properties and
their  finite-size   scaling  behavior.  Non-spanning  avalanches  and
3D-spanning avalanches exhibit mixed scaling behavior (indicated in
square brackets). They can be separated into two subcategories.}
\end{table*}

\begin{table}
\begin{tabular}{|c|c|c|c|}
\hline     
exponent      &     best     value   & Ref \\ 
\hline 
$\nu$ & $1.2 \pm 0.1$ & \onlinecite{PerezReche2003}\\
$\theta$ &  $0.10  \pm  0.02$  & \onlinecite{PerezReche2003}\\
$\theta_{nsc}$  & $2.02  \pm 0.04$  & \onlinecite{PerezReche2003} \\ 
$d_f$  & $2.78  \pm 0.05$  & \onlinecite{PerezReche2003},
this work \\
$d_{3-}$ & $2.98  \pm 0.02$ & \onlinecite{PerezReche2003}, this work \\
$\tau_{nsc}$ &  $1.65 \pm  0.02$ & \onlinecite{PerezReche2003} \\ 
$\beta_{c}$ &  $0.15\pm 0.08$  & \onlinecite{PerezReche2003} \\
$\beta_{3-}$ & $0.024\pm0.012$ &
\onlinecite{PerezReche2003} \\ 
$1/\mu$ & $ 1.5 \pm 0.1$ & this work \\
\hline
\end{tabular}
\caption{\label{TABLE2}  Summary   of  the  values   of  the  critical
exponents   of   the   3D-GRFIM   with   metastable   dynamics,
obtained in 
Ref. \onlinecite{PerezReche2003} and in the present work, as indicated
in the last column.}
\end{table}

Classification  of  the  avalanches  starts by  checking  whether  the
avalanches  span  the   system  in  1,2  or  3   spatial  directions.  
\footnote{For a  discussion of the exact definition  of a ``spanning''
  avalanche  see Ref.  \cite{PerezReche2003}.}   These three  kinds of
spanning avalanches are indicated by $\alpha=1,2,3$ respectively.

To  obtain good scaling  collapses of  the 3D-spanning  avalanches, an
extra  hypothesis  was introduced:  they  can  be  separated into  two
subcategories,  subcritical ($\alpha=3-$)  and  critical ($\alpha=3c$)
that scale  with different exponents.  In  particular, this assumption
indirectly  leads to  the  conclusion that  they  must have  different
fractal  dimensions   $d_{3-}$  and  $d_f$.    Moreover,  non-spanning
avalanches   ($\alpha=ns$)   should  also   be   separated  into   two
subcategories:    non    critical    ($\alpha=ns0$)    and    critical
($\alpha=nsc$),   depending   on  whether   their   number  and   size
distribution scales with distance to the critical point $u$ or not.

\begin{figure}[ht]
\begin{center}
\epsfig{file=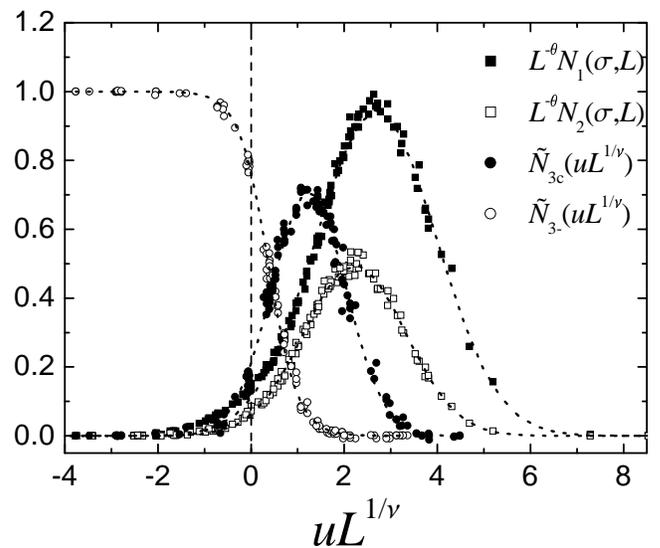, width=8.5cm}
\end{center}
\caption{\label{fig1} Scaling functions  corresponding to the number of
  1D-, 2D-, critical 3D- and subcritical 3D-spanning avalanches in the
  3D-RFIM as found in Ref. \cite{PerezReche2003}. The symbols show the
  overlap of the data corresponding to numerical simulations with many
  different system sizes ranging from $L=5$ to $L=48$.}
\end{figure}

Fig.~\ref{fig1}  presents a  summary \cite{Vives2003}  of  the scaling
functions  $\tilde{N}_{\alpha}$  according  to  the results  in  Ref.  
\cite{PerezReche2003}.   From the behavior  of such  scaling functions
when  $uL^{1/\nu}\rightarrow  \pm  \infty$  one  can  sketch  out  the
scenario in the thermodynamic limit.  Below $\sigma_c$ one subcritical
3D-spanning avalanche  exists, which  is responsible for  the observed
discontinuity of the magnetization in the thermodynamic limit.  Such a
discontinuity  is the  order parameter  and vanishes  when approaching
$\sigma_c^-$       according      to      $\Delta       m      \propto
(\sigma_c-\sigma)^{\beta_{3-}}$,        with        an        exponent
$\beta_{3-}=\nu(3-d_{3-})=0.024\pm0.012$.     Nevertheless,    it   is
difficult to  find this critical  behavior from simulations  of finite
systems  since  the contributions  from  critical spanning  avalanches
($\alpha=1$,$2$  and $3c$)  may  lead to  quite  good (but  incorrect)
scaling     collapses     of     the     order     parameter     using
$\beta_c=\nu(3-\theta-df)=0.15\pm0.08$.

A large  number of non-critical non-spanning avalanches  exist for the
whole  range of  $\sigma$.   They cannot  contribute  to any  observed
macroscopic  jump  since  their  size  is  vanishingly  small  in  the
thermodynamic  limit.   At   $\sigma=\sigma_c$  the  6  categories  of
avalanches  exist.  On  average, one  finds  $0.79\pm0.02$ subcritical
3D-spanning avalanches and an infinite  number of the five other types
of  avalanches. At the  critical point  (or close  enough to  it), the
distribution of avalanche sizes  is dominated by critical non-spanning
avalanches  and exhibits  an  approximate power-law  behavior with  an
exponent $\tau_{eff}=\tau_{nsc}+(3-\theta_{nsc})/d_f  = 2.00\pm 0.06$. 
This  power law  behavior is  restricted to  the central  part  of the
avalanche size distribution $1 \ll s \ll L^3$.

In the present  paper we will concentrate on the  4 different kinds of
spanning  avalanches, extending the  finite-size scaling  analysis and
focusing on  (i) the  study of  the values of  the external  field for
which the spanning avalanches occur and (ii) the direct measurement of
the geometrical  properties of the  avalanches.  This will  enable the
critical  exponent $\mu$,  related to  the renormalization  group (RG)
flow along the field direction, to  be found and have a direct test of
the FSS  hypothesis that leads to  the separation of the  two kinds of
3D-spanning  avalanches  (critical  and  subcritical)  with  different
fractal dimensions: $d_f=2.78\pm0.05$ and $d_{3-}=2.98\pm 0.02$.

In section  \ref{Model} we summarize  the 3D-GRFIM and the  details of
our  numerical simulations.   In section  \ref{Results}  raw numerical
results are  presented. In section  \ref{Scaling} we discuss  the main
FSS hypothesis, as an  extension of those presented previously.  These
hypotheses  are checked  in section  \ref{Sec:collapses}.   In section
\ref{fractal}   we  determine  the   geometrical  properties   of  the
avalanches.  In  section \ref{Discussion} we  discuss the consequences
of  the present study  and, finally,  in section  \ref{Conclusions} we
summarize and conclude our findings.

\section{Model}
\label{Model}

The 3D-GRFIM is defined on a cubic lattice of size $L\times L\times L$
with periodic boundary conditions.  On each lattice site ($i=1, \dots,
L^3$)  there is  a spin  variable $S_i$  taking values  $\pm  1$.  The
Hamiltonian is:
\begin{equation}
\label{Hamiltonian}
{\cal  H}=-\sum_{i,j}^{n.n.}  S_i  S_j -\sum_{i=1}^{L^3}  h_i S_i  - H
\sum_{i=1}^{L^3} S_i \; \; ,
\end{equation}
where the first sum extends over all nearest-neighbor (n.n) pairs, $H$
is the  external applied field  and $h_i$ are quenched  random fields,
which  are independent  and are  distributed according  to  a Gaussian
probability density with zero mean and standard deviation $\sigma$.

The  equilibrium ground-state  ($T=0$)  of this  Hamiltonian has  been
recently studied  \cite{Middleton2002}. In this  work we focus  on the
metastable version  of the 3D-GRFIM  proposed for the analysis  of the
behavior at $T=0$ when the system is driven by the external field $H$.
For $H=+ \infty$  the state of the system which  minimizes $\cal H$ is
the state  with maximum magnetization $m=\sum_{i=1}^{L^3} S_i  / L^3 =
1$.   When the  external field  $H$ is  decreased, the  system evolves
following local relaxation dynamics.   The spins flip according to the
sign of the local field:
\begin{equation}
\label{Eq:3}
h_i + H + \sum_{j=1}^6 S_j \; \; ,
\end{equation}
where  the  sum  extends  over  the 6  nearest-neighboring  spins  of
$s_i$. Avalanches occur when a spin flip changes the sign of the local
field of  some of the neighbors.   This may start a  sequence of spin
flips which occur at a fixed  value of the external field $H$, until a
new stable  situation is reached.   $H$ is then decreased  again.  The
size of the  avalanche $s$ corresponds to the  number of spins flipped
until a new  stable situation is reached. Note  that the corresponding
magnetization change is $\Delta m = 2 s / L^3$.

The  numerical algorithm  we have  used is  the so-called  brute force
algorithm which  propagates one avalanche at a  time \cite{Kuntz1999}. 
We have  studied system sizes  ranging from $L=5 (L^3=125)$  to $L=180
(L^3=5832000)$.  The  measured properties  are always averaged  over a
large  number of realizations  of the  random field  configuration for
each  value of  $\sigma$, which  ranges between  more than  $10^4$ for
$L\le 80$ to $300$ for $L=180$.

We  have  recorded the  sequence  of  avalanche  sizes during  half  a
hysteresis loop, i.e.  decreasing $H$ from $+\infty$ to $-\infty$.  We
have determined  not only the  size $s$ of each  individual avalanche,
but also the field $H$  at which each avalanche occurs. The avalanches
have  been classified  as non-spanning,  1D-spanning,  2D-spanning and
3D-spanning as explained in Ref.  \cite{PerezReche2003}.

By performing statistics, we  obtain the average density of avalanches
for    each   type   occurring    within   an    interval   $(H,H+dH)$
\footnote{Typical values  of $\Delta H$  range from $\Delta H  = 0.05$
  for $L  = 8$ to $\Delta  H=0.005$ for $L  = 48$}. We will  call this
quantity the number density $n_{\alpha} (H;\sigma,L)$. It satisfies:
\begin{equation}
\label{Eq:4}
\int_{-\infty}^{\infty} dH \; n_{\alpha} (H; \sigma,L) = N_{\alpha} (\sigma,L)
\end{equation}
where $N_{\alpha} (\sigma,L)$ are  the average number of avalanches of
each kind \cite{PerezReche2003} defined in Table \ref{TABLE1}. We have
also  measured  the  bivariate  size distribution  ${\cal  D}_{\alpha}
(s,H;\sigma,L)$.  This probability density is normalized so that:
\begin{equation}
\label{Eq:5}
\sum_{s=1}^{L^3} \int_{-\infty}^{\infty}  dH \; {\cal D}_{\alpha} (s,H;\sigma,L)
= 1
\end{equation}
Note  that by  projecting ${\cal  D}_{\alpha}$ we  can obtain  the two
marginal distributions:
\begin{equation}
\label{Eq:6}
\sum_{s=1}^{L^3} {\cal D}_{\alpha} (s,H;\sigma,L) =
\frac{n_{\alpha}(H;\sigma,L)}{N_{\alpha}(\sigma,L)}
\end{equation}
and
\begin{equation}
\label{Eq:7}
\int_{-\infty}^{\infty}    dH   \;    {\cal D}_{\alpha}    (s,H;\sigma,L)   =
D_{\alpha}(s;\sigma,L)
\end{equation}
which  represent the probability  density of  finding an  avalanche of
type  $\alpha$  within  $(H,H+dH)$  and  the  probability  (integrated
distribution) that an avalanche of type $\alpha$ has a size $s$ (after
the half loop) respectively.

In the following sections bivariate distributions will be presented as
point  clouds and  the  marginal distributions  as histograms.   Point
clouds  provide  qualitative   understanding  of  the  distributions.  
Quantitative  analysis  is much  better  performed  from the  marginal
distributions and their moments. The average field where the different
types of avalanches occur will be particularly interesting:
\begin{equation}
\langle  H   \rangle_{\alpha}(\sigma,L)=\int_{-\infty}^{\infty}  dH  \; H
\frac{n_{\alpha}(H;\sigma,L)}{N_{\alpha}(\sigma,L)},
\label{fieldaverage}
\end{equation}
and its standard deviation:
\begin{equation}
\label{Eq:8.2}
\sigma_{\alpha}^H (\sigma,L)=\left( \langle H^2 \rangle_{\alpha} -
  \langle H \rangle_{\alpha}^2 \right) ^{1/2}.
\end{equation}

In  some  simulations we  have  also  performed  a sand  box  counting
analysis  \cite{Bunde1994} of  each individual  spanning  avalanche in
order to have a direct  determination of their fractal dimension.  The
method consists  of considering boxes  of linear size  $\ell$ (ranging
from $1 \times 1\times 1$, $3 \times 3\times 3$, $5\times 5 \times 5$,
...  up  to $(L-1) \times (L-1)  \times (L-1)$) centred  near the spin
that  triggered each spanning  avalanche. We  determine the  number of
sites that  belong to the avalanche  for each box.   By averaging over
all the avalanches of the same kind (occurring during a half loop) and
over many disorder realizations,  we obtain the mass $M_{\alpha} (\ell
;\sigma,L)$ as a function of the box length $\ell$.

\section{Numerical results}
\label{Results}

\begin{figure}[ht]
\begin{center}
\epsfig{file=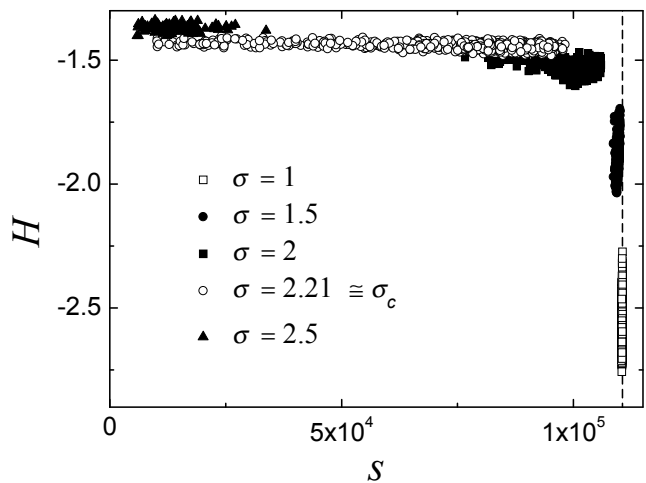, width=8.5cm}
\end{center}
\caption{\label{fig2}   Point  clouds  corresponding   to  3D-spanning
  avalanches for a system with size $L=48$.  Each point corresponds to
  an individual  avalanche and  indicates the size  $s$ and  the field
  where it  occurred $H$. Clouds corresponding to  different values of
  $\sigma$ are indicated by different symbols. Data have been averaged
  over many  realizations of disorder.  The dashed line  indicates the
  maximum value $s=L^3$.}
\end{figure}

Fig.~\ref{fig2}   shows  a   point  cloud   corresponding   to  ${\cal
  D}_{3}(s,H;\sigma,L)$ for  $L=48$ and different values  of $\sigma$. 
As  can be seen,  below $\sigma_c$  3D-spanning avalanches  are large,
close to  $s=L^3$ and exhibit a  certain spread around a  value of $H$
which  shifts upwards when  $\sigma_c$ is  approached from  below.  At
$\sigma=\sigma_c$ the  avalanches concentrate around  a critical field
value $H_c  \simeq -1.42$  and span almost  all possible  sizes. Above
$\sigma_c$ the few existing 3D-spanning avalanches are small. In fact,
they are smaller  than the mean size of a percolating  cluster in a 3D
cubic lattice ($s=0.311 L^3$).

\begin{figure}[ht]
\begin{center}
\epsfig{file=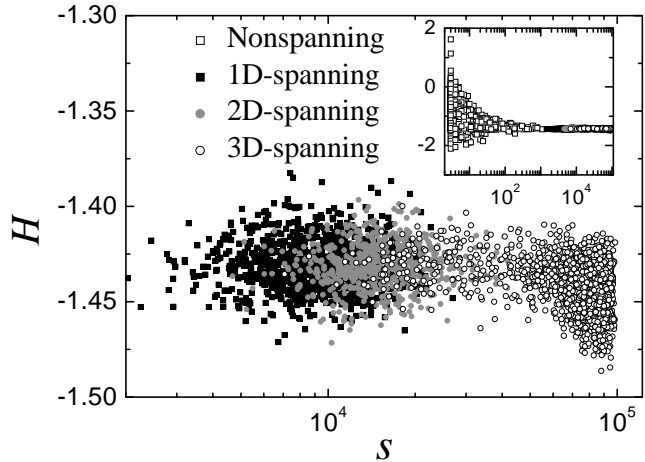, width=8.5cm}
\end{center}
\caption{\label{fig3}  Point   clouds  corresponding  to  1D-,
  2D-, and 3D-spanning avalanches at $\sigma=\sigma_c$ for $L=48$. The
  inset shows, on a log-linear  scale, the same data together with the
  cloud corresponding to non-spanning avalanches.}
\end{figure}

Fig.~\ref{fig3} shows the  point clouds corresponding to non-spanning,
1D-,   2D-,   and  3D-spanning   avalanches   at  $\sigma=\sigma_c$.   
Non-spanning  avalanches are suppressed  from the  main plot,  but are
shown  in the  inset on  a  log-linear scale.   As expected,  spanning
avalanches concentrate around  $H_c$. 3D-spanning avalanches exhibit a
larger size  than 2D-spanning  avalanches and the  latter show  a size
larger  than  1D-spanning  avalanches.   Moreover, one  can  see  that
3D-spanning avalanches are distributed in  a double cloud.  As will be
seen,  these two  clouds correspond  to the  two types  of 3D-spanning
avalanches predicted in \cite{PerezReche2003}.

\begin{figure}[ht]
\begin{center}
\epsfig{file=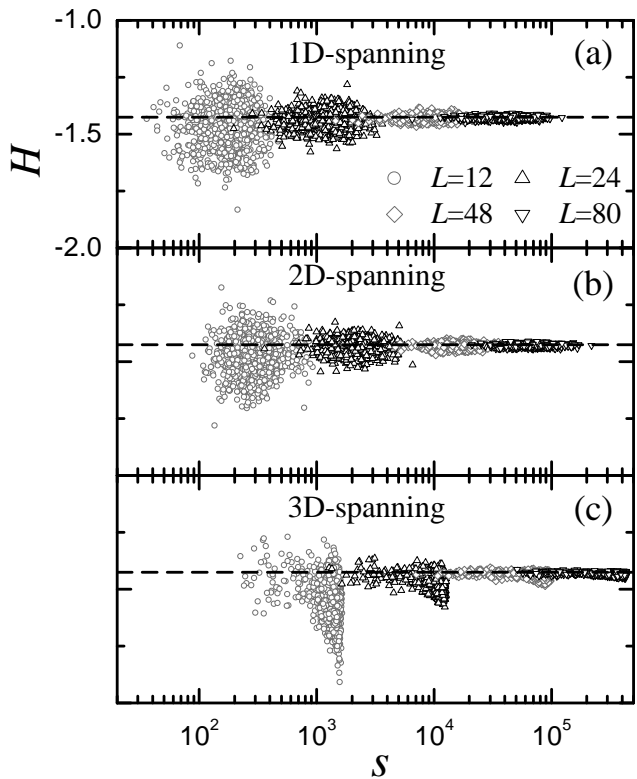, width=8.5cm}
\end{center}
\caption{\label{fig4} Point  clouds corresponding to 1D-  (a), 2D- (b)
  and 3D-spanning  avalanches (c) at  $\sigma=\sigma_c$ for increasing
  system sizes, as indicated by the  legend. The same scale is used on
  the three plots.}
\end{figure}

The   dependence  on   the   system  size   $L$   is  illustrated   in
Fig.~\ref{fig4}.   The  point  clouds  represent the  distribution  of
spanning avalanches at $\sigma=\sigma_c$ for increasing values of $L$.
The three  plots correspond to (a)  1D-, (b) 2D-,  and (c) 3D-spanning
avalanches.   The dashed line  corresponds to  our estimated  value of
$H_c=-1.425$.   Note that  all kinds  of spanning  avalanches  tend to
concentrate around  such a  field value for  increasing system  sizes. 
The shape of the cloud corresponding to 3D-spanning avalanches remains
assymetric, while $L$ grows.

\begin{figure}[ht]
\begin{center}
\epsfig{file=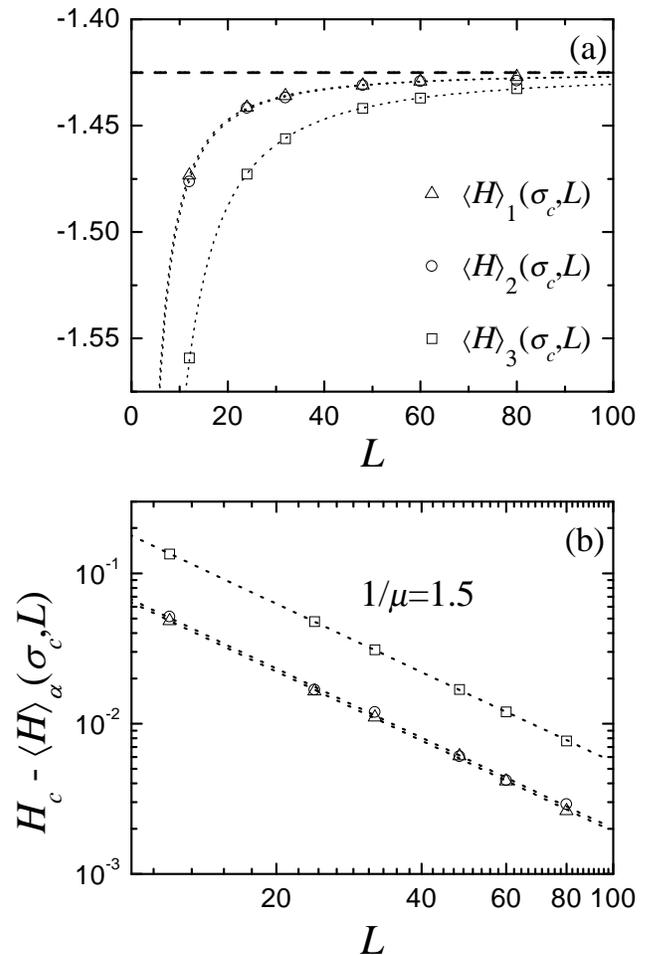, width=8.5cm}
\end{center}
\caption{\label{fig5} (a)  Average fields $\langle  H \rangle_{\alpha}
(\sigma_c,L)$.  The dashed lines indicates $H_c=-1.425$ and the dotted
lines correspond to  power-law fits. (b) The same  data represented on
log-log scales. The lines indicates the power-law behavior. Note that
the 3 kinds of spanning avalanches exhibit the same exponent.}
\end{figure}

To obtain a quantitative measure  of $H_c$, we have computed the field
averages $\langle  H \rangle_1$, $\langle H \rangle_2$  and $\langle H
\rangle_3$. Their  behavior as a function of  $L$ at $\sigma=\sigma_c$
is plotted in Fig.~\ref{fig5}(a). In  the next sections a FSS analysis
will  be  formally  proposed.   Nevertheless,  from  the  behavior  in
Fig.~\ref{fig5}(a)  one  can   already  guess  the  following  scaling
hypothesis:
\begin{equation}
\label{Eq:10}
\langle  H \rangle_{\alpha}(\sigma_c,L) - H_c \sim C_{\alpha} L^{1/\mu},
\end{equation}
where  $\mu$ will  be the  exponent  governing the  divergence of  the
correlation   length  when   the   field  $H$   approaches  $H_c$   at
$\sigma=\sigma_c$.  A  first check of this hypothesis  is performed in
Fig.~\ref{fig5}(b)  by a 3-parameter  ($H_c$, $\mu$  and $C_{\alpha}$)
least-squares fit  to the  three sets of  data.  We have  obtained the
same values  of $1/\mu= 1.5 \pm  0.1$ and $H_c=-1.425 \pm  0.010 $ for
$\langle  H   \rangle_1$,  $\langle   H  \rangle_2$  and   $\langle  H
\rangle_3$.  Results of  the fits  are  indicated by  dotted lines  in
Figs.~\ref{fig5}(a) and \ref{fig5}(b).

\section{Finite-Size Scaling hypothesis}
\label{Scaling}

\subsection{Scaling of distributions}
To proceed with the analysis  of the numerical data one must postulate
the ad-hoc FSS hypothesis.   As done in Ref.~\cite{PerezReche2003}, we
follow  standard RG  arguments.  The  distance to  the  critical point
$(\sigma_c,H_c)$ is measured  with two scaling variables $u(\sigma,H)$
and $v(\sigma,H)$,  which depend on the  externally tunable parameters
$\sigma$  and $H$. After  a renormalization  step that  transforms the
system size $L$ as:
\begin{equation}
\label{Eq:12.1}
L_b= b^{-1} L.
\end{equation}
$u$,$v$, and the size $s$ of an avalanche of type $\alpha$ will behave
as:
\begin{equation}
\label{Eq:12.2}
u_b= b^{1/\nu} u \; \; \; \; \; v_b= b^{1/\mu} v \; \; \; \; \; s_b= b^{-d_{\alpha}} s,
\end{equation}
where $d_\alpha$ are the fractal  dimensions of the avalanches and the
exponents $\nu$  and $\mu$ control  the divergence of  the correlation
length when  approaching the critical  point along the  two directions
$u$ and $v$.  Consequently, in  order to formulate the FSS hypothesis,
we will consider the following three main RG invariants:
\begin{equation}
\label{Eq:13.1}
u L^{1/\nu}=u_b L_b^{1/\nu} 
\end{equation}
\begin{equation}
\label{Eq:13.2}
v L^{1/\mu}=v_b L_b^{1/\mu} 
\end{equation}
\begin{equation}
\label{Eq:13.3}
s L^{-d_{\alpha}}=s_b L_b^{-d_{\alpha}} 
\end{equation}
Therefore,  close   enough  to   the  critical  point   the  bivariate
distributions ${\cal D}_{\alpha}(s,H;\sigma,L)$ will behave as:
\begin{equation}
\label{Eq:14}
{\cal D}_{\alpha}(s,H;\sigma,L)=L^{-{d_\alpha}+1/\mu}\hat{\cal{D}}_{\alpha}\left
  ( sL^{-d_{\alpha}},uL^{1/\nu},vL^{1/\mu} \right )  
\end{equation}
where  $d_\alpha$   is  $d_f=2.78$   for  the  1D-,2D-   and  critical
3D-spanning   avalanches  and   $d_{3-}=2.98$   for  the   subcritical
3D-spanning avalanches, as found in
Ref. \cite{PerezReche2003}.

The number densities $n_{\alpha}(H;\sigma,L)$ will behave as:
\begin{equation}
\label{Eq:15}
n_{\alpha}(H;\sigma,L)=L^{\theta_{\alpha}+1/\mu}   {\hat   n}_{\alpha}
\left ( uL^{1/\nu},vL^{1/\mu} \right ),
\end{equation}
where  $\theta_{\alpha}$  is  $\theta_c=0.1$  for  the  1D-,  2D-  and
critical  3D-spanning  avalanches  and  $\theta_{\alpha}= 0$  for  the
subcritical 3D-spanning  avalanches. The FSS  hypothesis (\ref{Eq:14})
and   (\ref{Eq:15})   extend  the   hypothesis   presented  in   Table
\ref{TABLE1}, including the dependence on the external field $H$.

\subsection{Detailed scaling variables}
The  proper dependence of  the scaling  variables $u$  and $v$  on the
tunable model parameters $\sigma$ and $H$ is unknown, but it should be
analytic \cite{Ma1973,Cardy1996}.   Keeping this in  mind, we consider
the  following  second-order  expansion around  $\sigma=\sigma_c$  and
$H=H_c$:
\begin{eqnarray}
\label{Eq:15.1}
u&=&u_1+Au_1^2+A'v_1+A''u_1v_1+A'''v_1^2\\
\label{Eq:15.2}
v&=&v_1+Bv_1^2+B'u_1+B''u_1v_1+B'''u_1^2,
\end{eqnarray}
where  $u_1=(\sigma-\sigma_c)/\sigma_c$ and $v_1=(H-H_c)/H_c$  are the
first-order scaling variables.  These variables represent the simplest
choice to measure the distance to  the critical point.  Now we want to
check  which  corrections to  the  first-order  scaling variables  are
really  important for large system sizes. Consider a function
$F(H;\sigma,L)$ that can be written as
\begin{equation}
\label{Eq:15.3}
F(H;\sigma,L)=L^a \hat{F}(uL^{1/\nu},vL^{1/\mu}),
\end{equation}
when the  correct scaling  variables are considered.   In the  case in
which we  try to obtain  scaling collapses with two  incorrect scaling
variables $u'$  and $v'$,  an explicit dependence  on $L$  may appear,
which makes  the collapse $L^{-a}  F$ impossible for  different system
sizes. In such a situation:
\begin{equation}
\label{Eq:15.4}
F(H;\sigma,L)=L^a \hat{G}(u'L^{1/\nu},v'L^{1/\mu},L),
\end{equation}
where we have introduced a ``non''-scaling function $\hat{G}$ that depends
explicitly  on   $L$.  By  construction,   $\hat{G}$  coincides  with
$\hat{F}(uL^{1/\nu},vL^{1/\mu})$   when   $u'=u$   and  $v'=v$.    The
effective scaling variables $u'$ and $v'$ can be expanded up to second
order:
\begin{eqnarray}
\label{Eq:15.7}
u'&=&u_1+{\cal A}u_1^2+{\cal A}'v_1+{\cal A}''u_1v_1+{\cal A}'''v_1^2\\
\label{Eq:15.8}
v'&=&v_1+{\cal B}v_1^2+{\cal B}'u_1+{\cal B}''u_1v_1+{\cal B}'''u_1^2.
\end{eqnarray}
Up   to    second   order,    the    difference   between
$\hat{G}(u'L^{1/\nu},v'L^{1/\mu},L)$                                and
$\hat{F}(uL^{1/\nu},vL^{1/\mu})$ is
\begin{eqnarray}
\nonumber
& &\hat{G}(u'L^{1/\nu},v'L^{1/\mu},L)-\hat{F}(uL^{1/\nu},vL^{1/\mu})=\\
\nonumber
& &L^{1/\nu} \hat{F}_1 (uL^{1/\nu},vL^{1/\mu}) (u'-u)\\
\nonumber
&+&L^{1/\mu}\hat{F}_2 (uL^{1/\nu},vL^{1/\mu}) (v'-v)\\
\nonumber
&+& \frac{1}{2}L^{2/\nu} \hat{F}_{11}(uL^{1/\nu},vL^{1/\mu})
(u'-u)^2\\
\nonumber
&+& \frac{1}{2} L^{2/\mu} \hat{F}_{22}(uL^{1/\nu},vL^{1/\mu})
(v'-v)^2\\
\label{Eq:15.9}
& + & L^{1/\nu+1/\mu} \hat{F}_{12}(uL^{1/\nu},vL^{1/\mu}) (u'-u)(v'-v),
\end{eqnarray}
where the subindex 1 and 2 in the scaling functions $\hat{F}$ indicate
the partial derivatives with respect to $uL^{1/\nu}$ and $vL^{1/\mu}$,
respectively.     Introducing    the    expansions    (\ref{Eq:15.1}),
(\ref{Eq:15.2}),     (\ref{Eq:15.7}),    and     (\ref{Eq:15.8})    in
Eq.~(\ref{Eq:15.9}), defining  an appropriate set  of eleven functions
$\hat{f}_i(u_1L^{1/\nu},v_1L^{1/\mu})$   from   the   derivatives   of
$\hat{F}$, and re-ordering the terms in (\ref{Eq:15.9}) we can write:
\begin{widetext}
\begin{eqnarray}
\nonumber
& &\hat{G}(u'L^{1/\nu},v'L^{1/\mu},L)-\hat{F}(uL^{1/\nu},vL^{1/\mu})=
({\cal A}'-A')L^{1/\nu-1/\mu} \left( \frac{v_1}{v}\right) \hat{f}_1
+({\cal B}'-B')L^{1/\mu-1/\nu} \left( \frac{u_1}{u}\right) \hat{f}_2\\
\nonumber
&+&({\cal A}-A)L^{-1/\nu} \left( \frac{u_1}{u}\right)^2 \hat{f}_3
+({\cal B}-B)L^{-1/\mu} \left( \frac{v_1}{v}\right)^2 \hat{f}_4\\
\nonumber
&+&({\cal A}''-A'')L^{-1/\mu} \left( \frac{u_1}{u}\right) \left(
  \frac{v_1}{v}\right) \hat{f}_5
+({\cal B}''-B'')L^{-1/\nu} \left( \frac{u_1}{u}\right) \left(
  \frac{v_1}{v}\right) \hat{f}_6\\
\nonumber
&+&({\cal A}'''-A''')L^{1/\nu-2/\mu} \left(\frac{v_1}{v}\right)^2
\hat{f}_7
+({\cal B}'''-B''')L^{1/\mu-2/\nu} \left(\frac{u_1}{u}\right)^2
\hat{f}_8\\
\nonumber
&+& ({\cal A}'-A')^2 L^{2/\nu-2/\mu} \left( \frac{v_1}{v}\right)^2
\hat{f}_9
+({\cal B}'-B')^2 L^{2/\mu-2/\nu} \left( \frac{u_1}{u}\right)^2
\hat{f}_{10}\\
\label{Eq:15.10}
&+&({\cal A}'-A')({\cal B}'-B') \left( \frac{u_1}{u}\right) \left(
  \frac{v_1}{v}\right) \hat{f}_{11},
\end{eqnarray}
\end{widetext}
where the  dependence of  the functions $\hat{f}_i$  on $u_1L^{1/\nu}$
and $v_1L^{1/\mu}$ is not written for simplicity.

Given  that  $1/\nu  =   0.8$  (Table  \ref{TABLE2})  and  $1/\mu=1.5$
(Fig.~\ref{fig5}), and noting that only the terms multiplying positive
powers  of  $L$ will  be  important  in  the thermodynamic  limit,  we
conclude that only terms in  which ${\cal B}'-B'$ appears represent an
explicit  dependence of  $\hat{G}(u'L^{1/\nu},v'L^{1/\mu},L)$ on  $L$. 
Therefore, only the  $B'$ term is relevant in  the thermodynamic limit
and  thus must  be considered  in the  expansions  (\ref{Eq:15.7}) and
(\ref{Eq:15.8}).   The remaining  coefficients may  be  neglected.  In
particular,  a term  proportional to  $u_1^2$ is  not necessary  if we
consider large system sizes.  Such an irrelevance was in fact observed
in  Fig.   8  of  Ref.~\cite{PerezReche2003}.  Nevertheless,  we  will
retain  the   quadratic  correction  $Au_1^2$  in   order  to  compare
appropriately   with  previous   results   \cite{PerezReche2003}.   In
summary,  we will  use the  following approximations  for  the scaling
variables:
\begin{eqnarray}
\label{Eq:15.5}
u&=&u_1+Au_1^2,\\
\label{Eq:15.6}
v&=&v_1+B'u_1.
\end{eqnarray}
The correction  (\ref{Eq:15.6}) associated with the  distance to $H_c$
was first introduced in  Ref.  \cite{Perkovic1996} where the parameter
analogous  to $B'$  was called  the ``tilting''  constant.   We should
mention  that  the  authors  demonstrate  the  importance  of  such  a
correction with  arguments that are slightly  different those proposed
here.

\subsection{Scaling of the field averages and standard deviations}
One is now  ready to deduce the scaling  behavior corresponding to the
field averages defined in Eq.~(\ref{fieldaverage}) and to the standard
deviations   defined  in   Eq.~(\ref{Eq:8.2}).   On   the   one  hand,
multiplying the marginal  distribution $n_{\alpha}/N_{\alpha}$ by $H$,
integrating  over   the  full  $H$  range,  and   using  the  relation
(\ref{Eq:15.6}), we find
\begin{equation}
\label{Eq:16}
H_c \left(1-B'u_1
\right) - \langle  H \rangle_{\alpha}(\sigma,L) =  L^{-1/\mu}{\hat h}_{\alpha}
\left (uL^{1/\nu} \right ).
\end{equation}
It is  useful to  define an ``effective''  disorder-dependent critical
field $H^{\ast}_c(\sigma)$ as:
\begin{equation}
\label{Eq:17}
H^{\ast}_c(\sigma)=H_c \left(1-B'u_1\right).
\end{equation}

For $\sigma=\sigma_c$  we recover  the scaling hypothesis  proposed in
Eq.~(\ref{Eq:10}) with $C_{\alpha}=\hat{h}_{\alpha}(0)$.  In this case
we obtain  an estimate  of $1/\mu$ that  is unaffected by  the tilting
constant $B'$.

On the other  hand, by performing similar calculations,  it is easy to
write $\sigma_{\alpha}^H(\sigma,L)$ as:
\begin{equation}
\label{Eq:17.2}
\sigma_{\alpha}^H(\sigma,L)=L^{-1/\mu} \hat{\sigma}_{\alpha}^H (uL^{1/\nu}).
\end{equation}
Notice that this scaling expression  is also unaffected by the tilting
constant.

\subsection{Separation of the two kinds of 3D-spanning avalanches}
\label{Separation}

From  the  FSS  analysis  of  the  integrated  distributions,  it  was
suggested in \onlinecite{PerezReche2003} that two kinds of 3D-spanning
avalanches exist  with different fractal  dimensions.  This assumption
allowed for  excellent collapses  of the scaling  plots. Nevertheless,
the separation of the scaling functions corresponding to the two kinds
of avalanches was  possible by using a double  FSS technique involving
the collapse of  data corresponding to three or  more different system
sizes.   The  propagation  of   the  statistical  errors  within  such
complicated  computations rendered  large  error bars  in the  scaling
functions and exponents.

Given the two  different fractal dimensions, it would  be desirable to
be  able to  perform  a direct  classification  of the  $3-$ and  $3c$
avalanches  during simulations. Nevertheless,  this desirable  idea is
not  possible since, in  a finite  system, a  good determination  of a
fractal dimension  is only  possible after performing  statistics of
many avalanches of the same kind.

In this  work we propose  two separation methods that,  although being
approximate (a  small fraction of avalanches are  not well classified)
give   enough  bias  to   the  statistical   analysis  to   allow  for
determination of the different  properties of subcritical and critical
3D-spanning avalanches.

The idea behind the methods is  that for a finite system that is below
$\sigma_c$, one basically finds one subcritical 3D-spanning avalanche.
The  other  types of  spanning  avalanches  may  occur only  close  to
$\sigma_c$.   Moreover, given  their different  fractal  dimension, we
expect  subcritical 3D-spanning  avalanches  to be  larger.  Thus,  we
propose the  following two methods,  which will be applied  only below
$\sigma_c$.

{\em Method  1}: The  larger 3D-spanning avalanche  in a half  loop is
classified as  subcritical.  The other 3D-spanning  avalanches will be
considered critical 3D-spanning avalanches.  (We have checked that the
larger 3D-spanning  avalanche is  also the last  3D-spanning avalanche
found  when decreasing the  field from  $H=+\infty$ to  $H=-\infty$ in
almost all the studied cases.)

{\em  Method 2}: We  classify a  3D-spanning avalanche  as subcritical
only when no other spanning avalanches occur during the half loop.  If
other  spanning  avalanches  occur,   we  classify  them  as  critical
3D-spanning.  The  idea behind this  method (which we will  discuss in
section~\ref{Discussion})  is  the  conjecture  that  the  subcritical
3D-spanning avalanche,  close to, but  below $\sigma_c$ fills  a large
fraction of the system and does not allow other spanning avalanches to
exist.

Table  \ref{TABLE3}  shows how  the  two  methods  classify a  certain
3D-spanning  avalanche   depending  on  whether   the  other  spanning
avalanches found in the half loop are 1D-, 2D- or 3D-spanning. In this
latter case the fact that the other 3D-spanning avalanche(s) found are
smaller or  larger than the  avalanche being classified must  be taken
into account.   The two methods only  differ in the case  in which the
3D-spanning  avalanche  being classified  is  the  largest, but  other
spanning avalanches (either 1D, 2D or smaller 3D) exist in the loop.

\begin{table}
\begin{tabular}{|c|c|c|c|}
\hline     
\multicolumn{1}{|p{3cm}|}{1D, 2D or smaller 3D-spanning avalanche
  exist} & 
\multicolumn{1}{p{2cm}|}{larger 3D-spanning avalanche exist}& Method 1 & Method 2 \\ 
\hline 
no & no & 3- & 3- \\
yes & no & 3- & 3c \\
no & yes & 3c & 3c \\
yes & yes & 3c & 3c \\
\hline
\end{tabular}
\caption{\label{TABLE3}  Summary  of  the  classification of  a  given
3D-spanning avalanche  according to the  two methods proposed  in the
text.}
\end{table}

\begin{figure}[ht]
\begin{center}
\epsfig{file=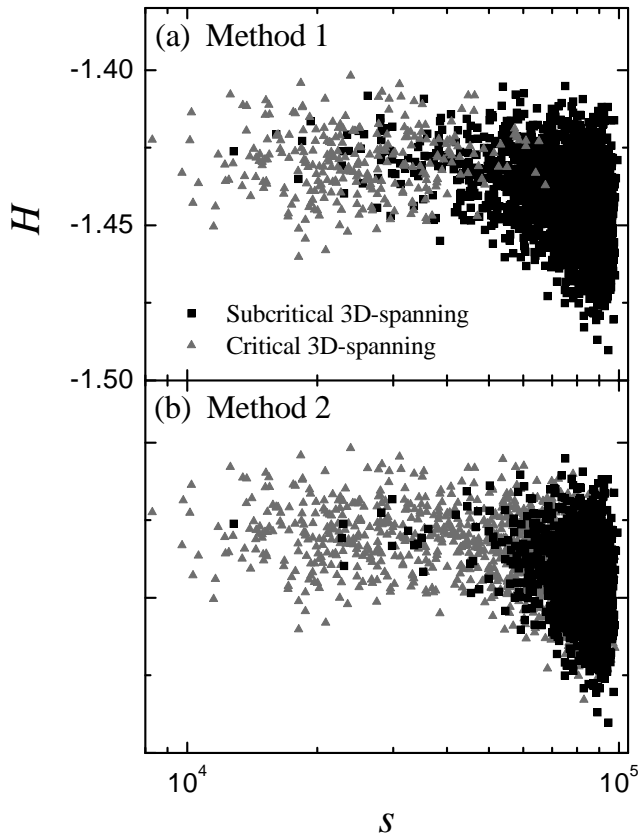, width=8.5cm}
\end{center}
\caption{\label{fig6} Example of  separation of 3D-spanning avalanches
  into subcritical  and critical, using  methods 1 and 2  explained in
  the text.}
\end{figure}

Fig.~\ref{fig6}  shows   an  example  of   separation  of  3D-spanning
avalanches into  subcritical and critical, using the  two methods.  It
corresponds to  $L=48$ and  $\sigma_c=2.21$.  One can  appreciate that
the  original double-shaped cloud  is separated  into two.   The cloud
corresponding to  critical 3D-spanning avalanches is  similar in shape
to the  clouds corresponding to 1D- and  2D-spanning avalanches.  Note
that  Method  2  classifies  a  certain number  of  large  avalanches,
occurring at  very negative  fields, as being  critical that  Method 1
classifies as being subcritical.

The two  separation methods  will be used  throughout the rest  of the
text  to  separately analyze  the  data  corresponding to  subcritical
3D-spanning avalanches  and critical 3D-spanning  avalanches.  In some
of the statistical analysis presented  below we will consider only the
3D-spanning avalanches which are equally classified by the two methods
and discard those which are  classified differently from the analysis. 
Although this procedure reduces the size of the statistical sample, it
ensures that we do not introduce any bias due to ill-classification of
some of the avalanches.

\section{Scaling collapses}
\label{Sec:collapses}
\subsection{Field Averages and standard deviations}
\label{Sec:FieldAverages} 
\begin{figure}[ht]
\begin{center}
\epsfig{file=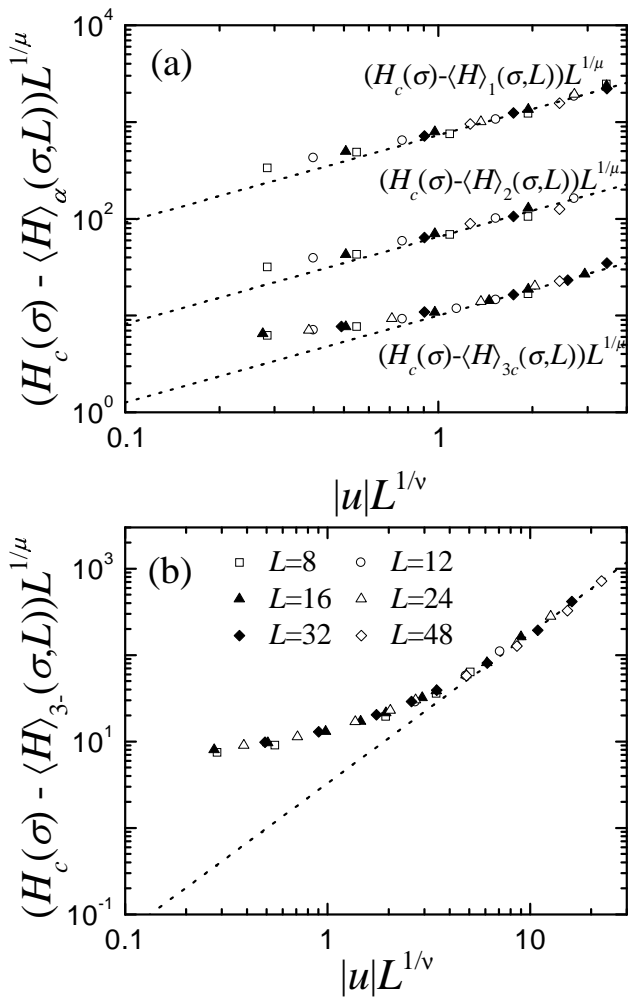, width=8.5cm}
\end{center}
\caption{\label{fig7} Scaling plots corresponding to the difference
  between  $\langle  H \rangle_{\alpha}$  and  the effective  critical
  disorder $H_c(\sigma)$  for (a)  1D-, 2D-, and  critical 3D-spanning
  avalanches  and  (b)  the  subcritical 3D-spanning  avalanches.   We
  present only data for $\sigma<\sigma_c$. The dotted lines correspond
  in each case to  the asymptotic behavior of $-\hat{h}_{\alpha}$ when
  $|u|L^{1/\nu} \rightarrow \infty$.   The scaling plots corresponding
  to the 1D- and 2D-spanning avalanches have been displaced one decade
  upwards for clarity.}
\end{figure}

Fig.~\ref{fig7}  presents the scaling  collapses corresponding  to the
field averages for $\sigma \le \sigma_c$. Data is presented on log-log
scales in  order to analyze  the power-law behavior  for $|u|L^{1/\nu}
\rightarrow  \infty$. Fig.~\ref{fig7}(a)  shows data  corresponding to
$\langle  H   \rangle_1$,  $\langle  H  \rangle_2$,   and  $\langle  H
\rangle_{3c}$, whereas Fig.~\ref{fig7}(b)  shows data corresponding to
$\langle  H  \rangle_{3-}$.   We  remark  that  only  the  3D-spanning
avalanches equally  classified by Methods 1  and 2 have  been used for
computing the  averages corresponding to $\langle  H \rangle_{3-}$ and
$\langle H \rangle_{3c}$.  By  fixing $1/\mu=1.5$ and taking $\nu=1.2$
from Ref.~\cite{PerezReche2003} we get the best collapses for $B'=0.25
\pm 0.10$.   We would  like to  emphasize that the  four sets  of data
scale extremely well with the same values of $\mu$ and $\nu$.

The  asymptotic behavior  of  ${\hat  h}_{3-}$ (dotted  line  in Fig.  
\ref{fig7}(b)) for  large values  of $|u|L^{1/\nu}$ is  $3.3\left( |u|
  L^{1/\nu}  \right)^{1.8}$.   The  exponent  $1.8$  equals  $\nu/\mu$
within  statistical  error.  This  means  that,  in the  thermodynamic
limit,
\begin{equation}
\label{Eq:18.1}
\langle H \rangle_{3-} (\sigma)= H_c^{\ast}(\sigma)- 3.3|u|^{\nu/\mu}.
\end{equation} 

\begin{figure}[ht]
\begin{center}
\epsfig{file=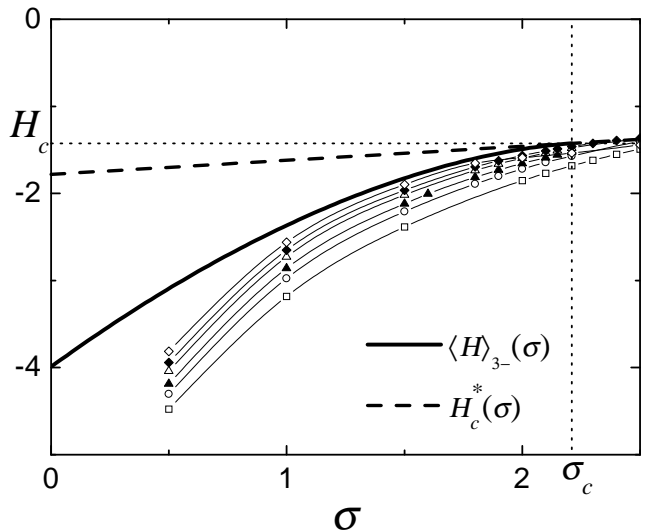, width=8.5cm}
\end{center}
\caption{\label{fig7.2} Representation of $\langle H
  \rangle_{3-}(\sigma)$   (Eq.~(\ref{Eq:18.1})),  $H^{\ast}_c(\sigma)$
  (Eq.~(\ref{Eq:17})), and $H_c$.  Symbols correspond to the numerical
  estimate of $\langle H  \rangle_{3-}$ from simulations for different
  $L$, as indicated by the legend in the previous figure.}
\end{figure}

Fig.~\ref{fig7.2} shows this behavior,  which finishes at the critical
point $(\sigma_c,H_c)$  because no subcritical  3D-spanning avalanches
exist     above.      The     disorder-dependent    critical     field
$H^{\ast}_c(\sigma)$  and the  critical field  $H_c$ are  indicated by
dashed  and dotted  lines,  respectively.  We  have  also plotted  the
numerical estimates  of $\langle H \rangle_{3-}$  for different system
sizes  in  order to  show  that  Eq.~(\ref{Eq:18.1})  is the  limiting
behavior for $L \rightarrow \infty$.
The  asymptotic behavior of  $\hat{h}_{\alpha}$ for  the 1D-,  2D- and
critical 3D-spanning avalanches is proportional to $\left(|u|L^{1/\nu}
\right)^{0.9}$. This implies that in the thermodynamic limit
\begin{equation}
\label{Eq:18.2}
\langle H \rangle_{\alpha}(\sigma) = H^{\ast}_c(\sigma)
\end{equation}
for  the   1D-,  2D-  and  critical  3D-spanning   avalanches.

Similar  finite-size scaling  analyses can  be done  for  the standard
deviations  of   the  marginal  distributions  $n_{\alpha}/N_{\alpha}$
according  to  Eq.~(\ref{Eq:17.2}).  From  the  obtained collapses  we
deduce  the  following behavior  for  large  values of  $|u|L^{1/\nu}$
($\sigma   <   \sigma_c$):   $\hat{\sigma}_{3-}^{H}(uL^{1/\nu})   \sim
(|u|L^{1/\nu})^{0.6}$ for  the subcritical 3D-spanning  avalanches and
$\hat{\sigma}_{3c}^{H}(uL^{1/\nu}) \sim  (|u|L^{1/\nu})^{0.2}$ for the
1D-,  2D-, and  critical 3D-spanning  avalanches. Similar  behavior is
observed    for   $\sigma   >    \sigma_c$.    These    results   (see
Eq.~(\ref{Eq:17.2}))  indicate  that  the  standard deviation  of  the
marginal  distribution  $n_{\alpha}/N_{\alpha}$  corresponding to  any
type of spanning avalanche vanishes in the thermodynamic limit for any
value of $\sigma$.

\subsection{Number density}
\label{Numberdens}

\begin{figure}[ht]
\begin{center}
\epsfig{file=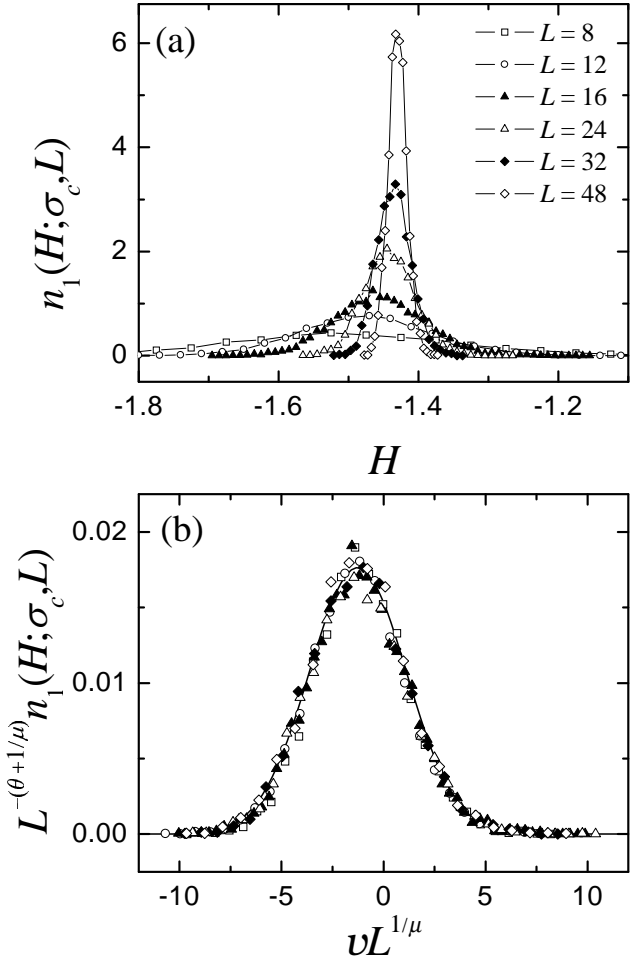, width=8.5cm}
\end{center}
\caption{\label{fig8} (a) Number density of spanning avalanches in one
  dimension  at the  critical amount  of disorder.   (b)  Scaling plot
  corresponding to the data in (a) according to Eq.~(\ref{Eq:15}) with
  $\theta=0.1$. The continuous line in (b) shows a Gaussian fit.}
 \end{figure}

\begin{figure}[ht]
\begin{center}
\epsfig{file=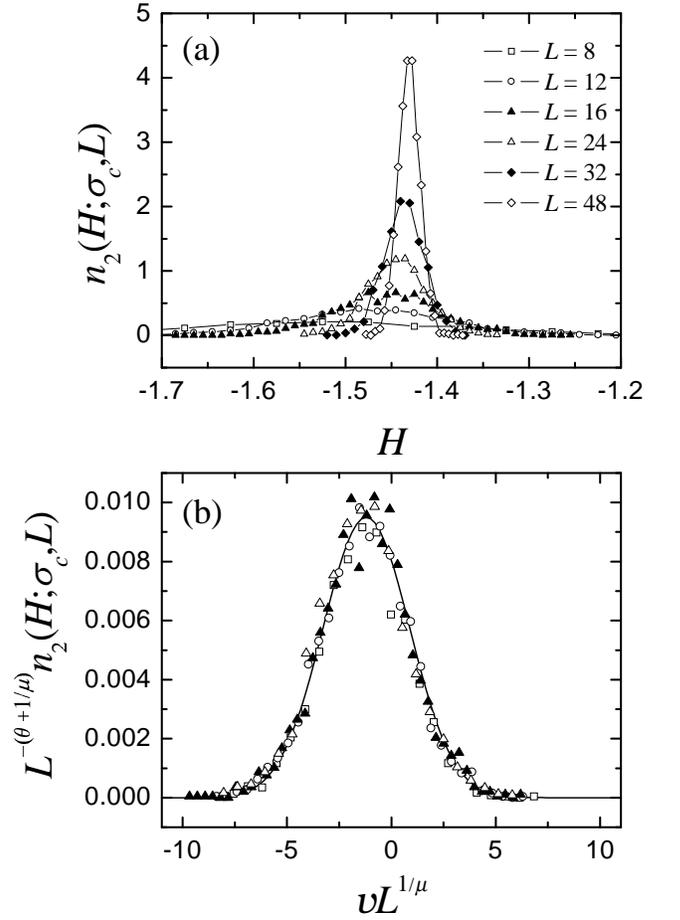, width=8.5cm}
\end{center}
\caption{\label{fig9} (a) Number density of spanning avalanches in two
  dimensions  at the critical  amount of  disorder.  (b)  Scaling plot
  corresponding to the data in (a) according to Eq.~(\ref{Eq:15}) with
  $\theta=0.1$. The continuous line in (b) shows a Gaussian fit.}
\end{figure}

The number density corresponding  to the 1D-spanning avalanches at the
critical  amount of  disorder  $n_1(H;\sigma_c,L)$ is  shown  in Fig.  
\ref{fig8}(a) as a function of  the applied field for different system
sizes. The number  density shows a peak that  increases and shifts for
increasing $L$.  Similar  behavior is observed for $n_2(H;\sigma_c,L)$
(Fig.~\ref{fig9}(a)).  A  FSS analysis is performed  using the scaling
assumption for the  number densities (Eq.~(\ref{Eq:15})).  The results
of  such   an  analysis  are  presented   in  Figs.~\ref{fig8}(b)  and
\ref{fig9}(b)   for   $n_1(H;\sigma_c,L)$   and   $n_2(H;\sigma_c,L)$,
respectively.  To obtain these collapses we have used $\theta=0.1$ and
$1/\mu =1.5$  as in the  preceding sections. The scaling  functions in
Figs.~\ref{fig8}(b)  and   \ref{fig9}(b)  are  well   approximated  by
Gaussian functions (indicated  by continuous lines).  When $vL^{1/\mu}
\rightarrow  \pm \infty$  both scaling  functions go  exponentially to
zero.  This  behavior indicates that,  in the thermodynamic  limit for
$\sigma=\sigma_c$,  1D-  and  2D-spanning  avalanches  only  exist  at
$H=H_c$.

\begin{figure}[ht]
\begin{center}
\epsfig{file=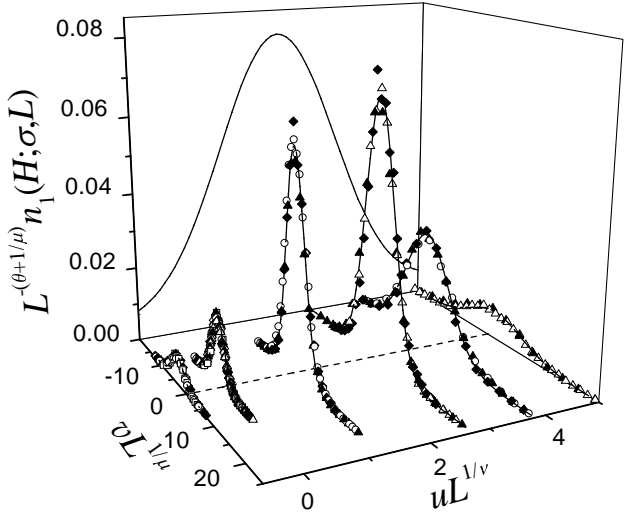, width=8.5cm}
\end{center}
\caption{\label{fig10} Collapses corresponding to
  $\hat{n}_1(uL^{1/\nu},vL^{1/\mu})$. The cuts  of the scaling surface
  are taken  at $uL^{1/\nu}=-0.58,0,1.22,2.7,3.8$ and  $5.0$.  Symbols
  correspond  to the  sizes indicated  in the  legend of  the previous
  figure.   The dashed  line  on the  horizontal  plane indicates  the
  tendency of the crest of the scaling surface.  The projection of the
  crest (Gaussian fit) is depicted with a continuous line on the plane
  $uL^{1/\nu}$-$\hat{n}_1$.}
\end{figure}

Fig.~\ref{fig10}  shows  several  cuts  corresponding to  the  scaling
surface  $\hat{n}_1(uL^{1/\nu},vL^{1/\mu})$.   From  the collapses  we
obtain $B'=0.25 \pm 0.10$ in  total agreement with previous estimates. 
From  a qualitative  point of  view, the  collapses indicate  that the
scaling   surface  shows   a   crest  with   amplitude  depending   on
$uL^{1/\nu}$.  More quantitatively, the scaling collapses for each cut
can be well approximated  by Gaussian functions, whose amplitude, peak
position,  and   width  depend  on   $uL^{1/\nu}$.   Furthermore,  the
dependence on $uL^{1/\nu}$ of  the fitted amplitudes also adjusts very
well  to a Gaussian  function that  follows the  profile of  the crest
(continuous line  on the back plane in  Fig.~\ref{fig10}).  The dashed
line on the bottom plane  indicates the position of the crest $\langle
vL^{1/\mu}  \rangle_{1}(uL^{1/\nu})=\hat{h}_1(uL^{1/\nu})/H_c$,  which
has already been shown in Fig.~\ref{fig7}(a).

All these considerations imply that, for any value of $H$, the scaling
function $\hat{n}_1$ decays exponentially when $uL^{1/\nu} \rightarrow
\pm  \infty$.   This  indicates  that,  in  the  thermodynamic  limit,
irrespective  of the  value of  $H$, $n_1$  is zero  for  $\sigma \neq
\sigma_c$.   In contrast,  when $\sigma=\sigma_c$,  $n_1$  diverges at
$H=H_c$ and is zero for other  values of the field.  This scenario for
$n_1$ is also applicable to $n_2$.

\begin{figure}[ht]
\begin{center}
\epsfig{file=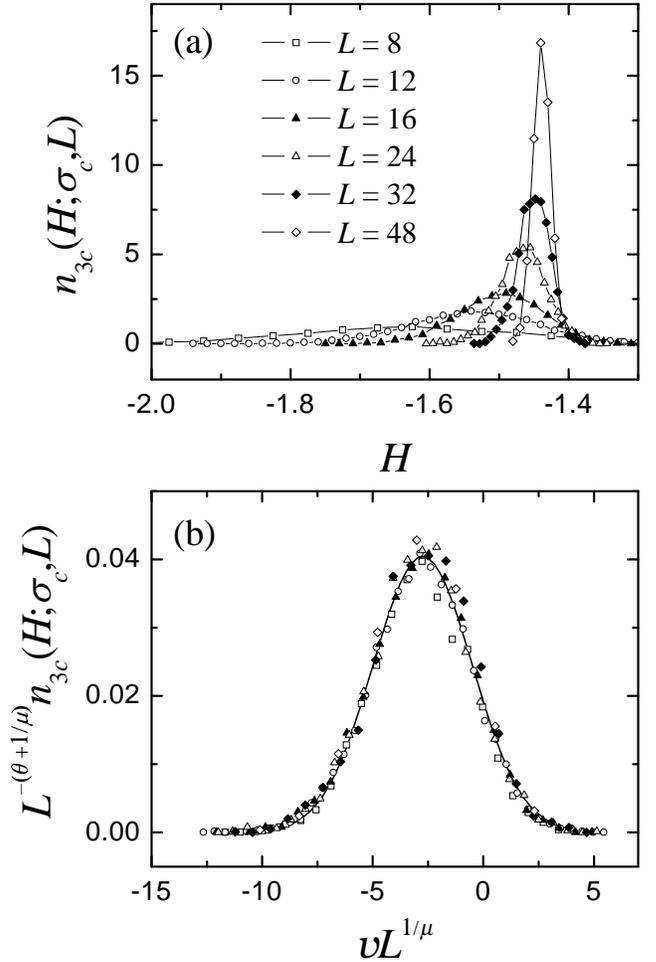, width=8.5cm}
\end{center}
\caption{\label{fig11} (a) Number density of critical 3D-spanning
  avalanches at the critical  amount of disorder obtained using Method
  2.  (b) Scaling plot  corresponding to the data in  (a) according to
  Eq.~(\ref{Eq:15}) with $\theta=0.1$.}
\end{figure}

\begin{figure}[ht]
\begin{center}
\epsfig{file=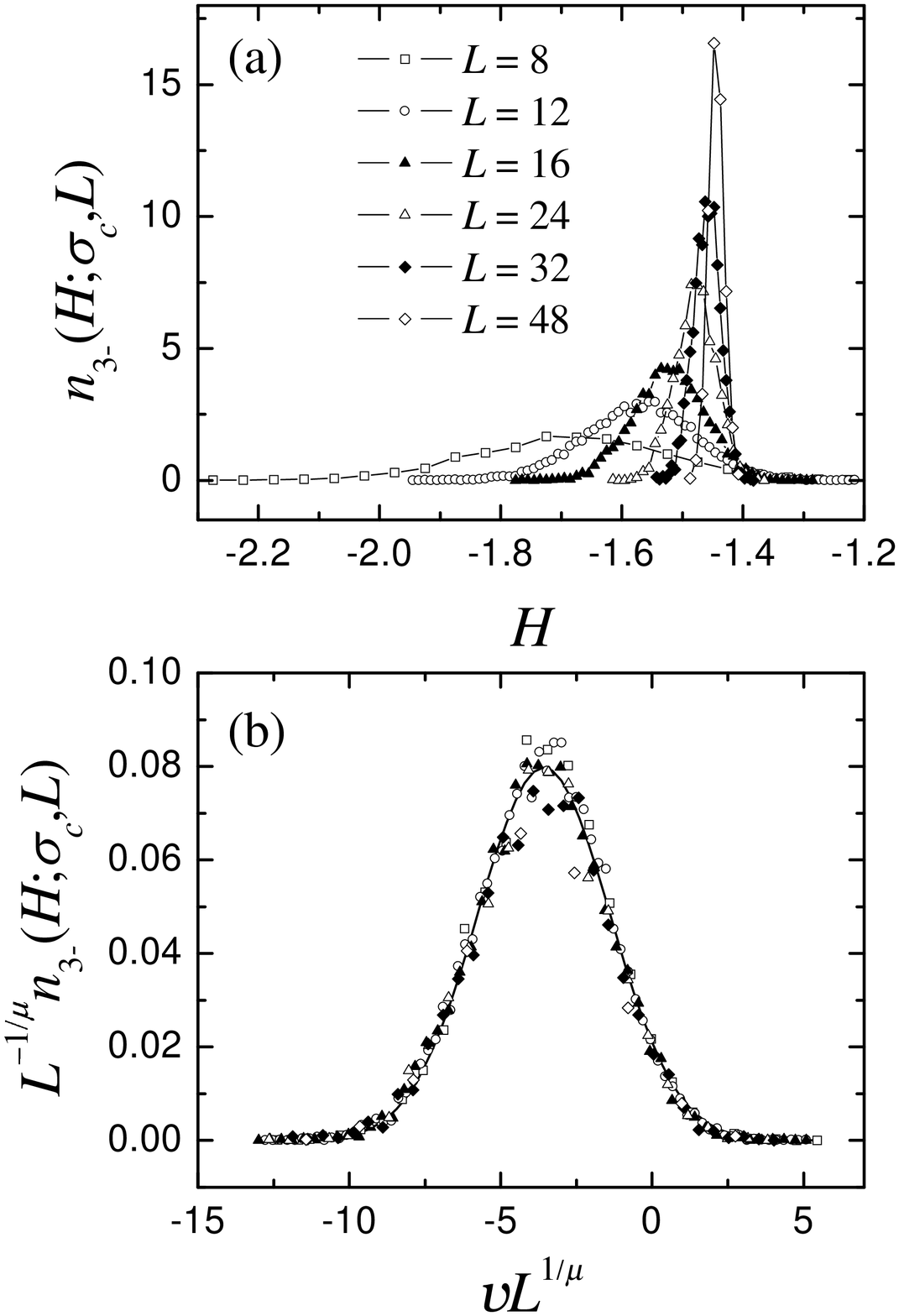, width=8.5cm}
\end{center}
\caption{\label{fig12} (a) Number density of subcritical 3D-spanning
  avalanches at the critical  amount of disorder obtained using Method
  2.  (b) Scaling plot  corresponding to the data in  (a) according to
  Eq.~(\ref{Eq:15}) with $\theta=0$.}
\end{figure}

To obtain  $n_{3c}$ and $n_{3-}$ we  have used Method  2 of separation
described in Sec.~\ref{Separation}.  The results for $\sigma=\sigma_c$
are    presented   in    Figs.~\ref{fig11}    and   \ref{fig12}    for
$n_{3c}(H;\sigma_c,L)$  and $n_{3-}(H;\sigma_c,L)$,  respectively.  We
have also tried  to separate the two kinds of  avalanches by Method 1,
but the  collapses are not  as good as  those obtained with Method  2. 
This  result  indicates that  in  the  set  of avalanches  non-equally
classified  by the two  methods, there  are more  critical 3D-spanning
avalanches than subcritical 3D-spanning avalanches.

As in the  case of $n_1$ and $n_2$  at $\sigma=\sigma_c$, the behavior
of  the scaling functions  in Figs.~\ref{fig11}(b)  and \ref{fig12}(b)
indicate that  both $n_{3c}$ and  $n_{3-}$ diverge at $H=H_c$  and are
zero for fields different to $H_c$.
The  detailed  study  of  the  bivariate  collapses  corresponding  to
$n_{3c}(H;\sigma,L)$   and  $n_{3-}(H;\sigma,L)$   for   $\sigma  \neq
\sigma_c$  is difficult  and  tedious. In  particular,  for $\sigma  >
\sigma_c$  we do not  expect the  separation methods  to work  and for
$\sigma < \sigma_c$, the analysis would require a lot of statistics.

\section{Direct determination of the fractal dimensions} 
\label{fractal}

In the  thermodynamic limit we  assume the standard  fractal behavior,
i.e.   that the  average mass  belonging to  a certain  avalanche type
inside a box of linear size $\ell$ is given by:
\begin{equation}
\label{Eq:30.1}
M_{\alpha}(\ell;\sigma) = M_{\alpha}^{\ast}(\sigma) \ell^{d_{\alpha}}
\end{equation}
in the  limit $\ell \ll \xi$,  where $\xi$ is the  correlation length. 
The prefactor $M_{\alpha}^{\ast}(\sigma)$ is related to the concept of
lacunarity  \cite{Mandelbrot1983}.  In general  there can  be fractals
sharing the  same fractal dimension, but with  different lacunarities. 
The fractal  dimension is related  to the rate  of change of  the mass
when the size  of the box is changed.  In  contrast, the lacunarity is
related to the  size of the gaps of the fractal  and is independent of
the fractal  dimension.  In this way,  the larger the  typical size of
the   gaps,   the  higher   the   lacunarity.    For  many   fractals,
\cite{Mandelbrot1983,Allain1991}   as    lacunarity   increases,   the
prefactor $M_{\alpha}^{\ast}$,  decreases since the mass  inside a box
of linear size $\ell$ decreases.

For   finite  systems   it   is  necessary   to   translate  the   law
(\ref{Eq:30.1})  into  a finite-size  scaling  hypothesis. As  usually
done, we propose:
\begin{equation}
\label{Eq:30.2}
M_{\alpha}(\ell;\sigma,L)=L^{d_{\alpha}}\hat{M}_{\alpha}(uL^{1/\nu},\ell/L)\;;
\hspace{0.7cm} \xi \lesssim L,
\end{equation}
where the condition $\xi \lesssim  L$ stands for the fact that scaling
only holds in the  critical zone.  Eq.~(\ref{Eq:30.2}) allows the data
corresponding to the masses $M_{\alpha}$  to be collapsed and, in this
way, to  obtain the fractal  dimensions $d_{\alpha}$.  We  can predict
the shape  of $\tilde{M}_{\alpha}$ in  two limiting cases: on  the one
hand,   the  scaling   function  $\hat{M}_{\alpha}(uL^{1/\nu},\ell/L)$
should behave as:
\begin{equation}
\label{Eq:30.3}
\hat{M}_{\alpha}(uL^{1/\nu},\ell/L) = M_{\alpha}^{\ast}(uL^{1/\nu})
\left( \frac{\ell}{L} \right)^{d_{\alpha}},
\end{equation}
in the limit  $\ell/L \ll \xi/L \lesssim 1$  to recover the expression
(\ref{Eq:30.1}) from (\ref{Eq:30.2}). On  the other hand, in the limit
$\ell/L  \gg  \xi/L$,  the   scaling  function  corresponding  to  the
subcritical 3D-spanning avalanche should behave as
\begin{equation}
\label{Eq:30.5}
\hat{M}_{3-}(uL^{1/\nu},\ell/L) \sim \left(|u|L^{1/\nu}\right)^{\beta_{3-}}
\left( \frac{\ell}{L}\right)^{3},
\end{equation}
if this  avalanche fills a  finite fraction $|u|^{\beta_{3-}}$  of the
system    for   $\sigma<\sigma_c$    in   the    thermodynamic   limit
\cite{Vives2003} and we expect
\begin{equation}
\label{Eq:30.4}
M_{3-}(\ell;\sigma) \sim |u|^{\beta_{3-}} \ell^{3}.
\end{equation}
Such   behavior  can   be  obtained   from   Eqs.~(\ref{Eq:30.5})  and
(\ref{Eq:30.2})       using       the      hyperscaling       relation
$\beta_{3-}=\nu(3-d_{3-})$.

\begin{figure}[ht]
\begin{center}
\epsfig{file=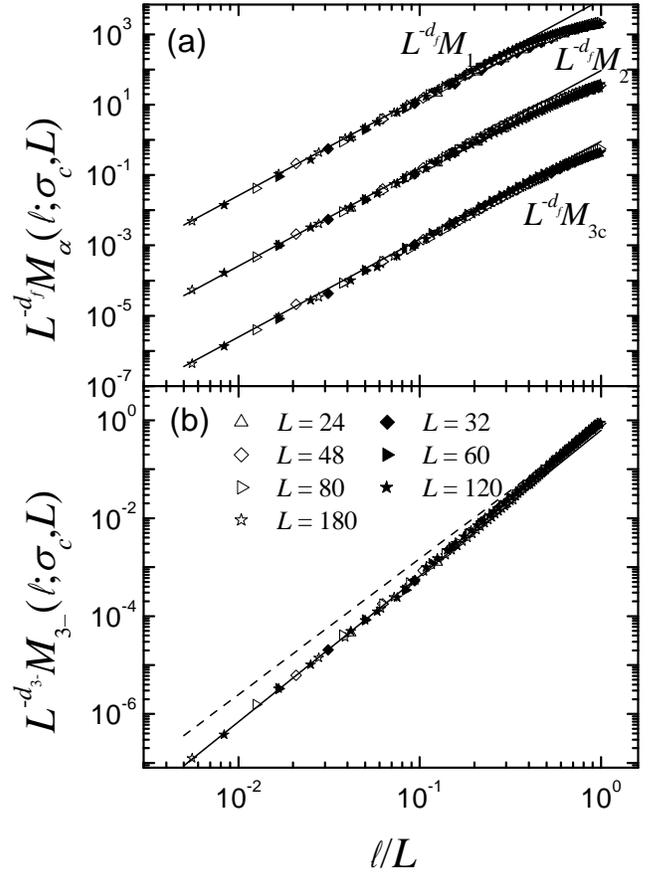, width=8.5cm}
\end{center}
\caption{\label{fig13} (a) Scaling collapses corresponding to the mass
  of   the  1D-,   2D-,   and  critical   3D-spanning  avalanches   at
  $\sigma=\sigma_c$.  The  collapses corresponding to  $M_1$ and $M_2$
  have been displaced two decades upwards for clarity.  (b) Scaling of
  $M_{3-}$  also at  $\sigma=\sigma_c$.  In  all cases  the asymptotic
  behavior  for small  values  of $\ell  /  L$ has  been indicated  by
  continuous  lines.   The  dashed  line  in (b)  corresponds  to  the
  assimptotic   behavior  of   $\hat{M}_{3c}$  to   compare   it  with
  $\hat{M}_{3-}$.}
\end{figure}

Fig.~\ref{fig13}  shows the  scaled  mass for  all  kinds of  spanning
avalanches from  simulations performed at  $\sigma=\sigma_c$.  We have
only  considered the  mass of  those 3D-spanning  avalanches  that are
equally classified by the two proposed methods.  The fractal dimension
rendering the best collapse is $d_{f}=2.78 \pm 0.05$ for the 1D-, 2D-,
and critical 3D-spanning avalanches and $d_{3-}=2.98 \pm 0.02$ for the
subcritical  3D-spanning   avalanches.   Both  values   are  in  total
agreement with those  obtained independently in \cite{PerezReche2003}. 
Moreover,  the  slope of  each  collapse in  the  limit  $\ell \ll  L$
(left-hand side of the collapses) coincides with the fractal dimension
used to obtain the collapses,  so that the behavior (\ref{Eq:30.3}) is
confirmed.  The  prefactors are $M_1^{\ast}(\sigma_c)=0.95  \pm 0.07$,
$M_2^{\ast}(\sigma_c)=0.93  \pm  0.07$,  $M_{3c}^{\ast}(\sigma_c)=0.90
\pm 0.07$, and $M_{3-}^{\ast}(\sigma_c)=0.65 \pm 0.07$.  The low value
of  the   prefactor  corresponding  to   the  subcritical  3D-spanning
avalanches indicates that the gaps  of these avalanches are large.  As
a consequence, the space filled  locally by these avalanches is not as
high  as  one would  \emph{a  priori}  think  given the  proximity  of
$d_{3-}$ to 3.

\begin{figure}[ht]
\begin{center}
\epsfig{file=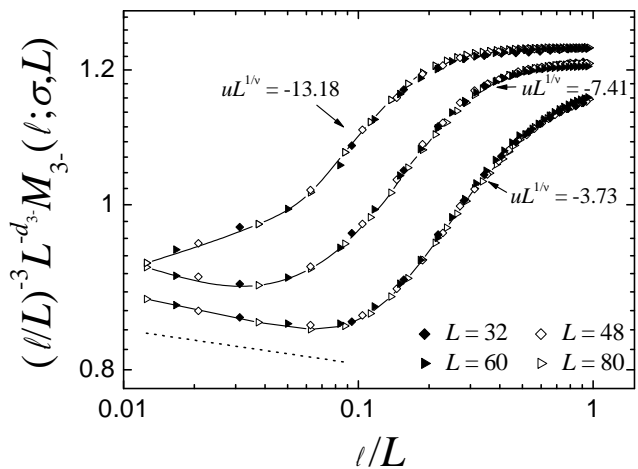, width=8.5cm}
\end{center}
\caption{\label{fig15} Collapses corresponding to
  $(\ell/L)^{-3}\tilde{M}_{3-}(uL^{1/\nu},\ell/L)$                  for
  $uL^{1/\nu}=-3.73,-7.41$ and $-13.18$. The dashed line indicates the
  slope $d_{3-}-3=0.02$. Continuous lines are a guide to the eye.}
\end{figure}

To  study  the  behavior  of  $M_{3-}$  for  $\sigma<\sigma_c$  it  is
convenient to  multiply the  scaling function $\tilde{M}_{3-}$  by the
factor $(\ell/L)^{-3}$. From Eqs.~(\ref{Eq:30.3}) and (\ref{Eq:30.5}),
it should behave as:
\begin{equation}
\label{Eq:31}
\left( \frac{\ell}{L} \right)^{-3} \tilde{M}_{3-}=
\begin{cases}
M^{\ast}_{3-}(uL^{1/\nu})\left(\frac{\ell}{L} \right)^{d_{3-}-3} &,
\frac{\ell}{L} \ll \frac{\xi}{L} \\ 
\sim \left( |u|L^{1/\nu} \right)^{\beta_{3-}} &, \frac{\ell}{L} \gg \frac{\xi}{L}
\end{cases}%
\end{equation}
in such a way that, for  a given value of $uL^{1/\nu}<0$, the function
$(\ell/L)^{-3}\tilde{M}_{3-}$  approaches a  constant value  for large
values   of  $\ell/L$   if   the  correlation   length   is  finite.   
Fig.~\ref{fig15}      shows     the      scaling      collapses     of
$(\ell/L)^{-3}\tilde{M}_{3-}$  for three cuts  of the  scaling surface
taken at  $uL^{1/\nu}=-3.73,-7.41$ and  $-13.18$. Note that  such cuts
are limited from below at  $\ell/L=1/80$. In spite of this limitation,
the results  clearly indicate that  (i) for small values  of $\ell/L$,
the  behavior of  $(\ell/L)^{-3}\tilde{M}_{3-}$ is  power law  with an
exponent approaching  $d_{3-}-3 \simeq 0.02$ (indicated  by the dotted
line)   and  (ii)   for  large   values  of   $\ell/L$   the  function
$(\ell/L)^{-3}\tilde{M}_{3-}$ tends to a constant value.  (This latter
tendency  can   only  be  observed  for  negative   enough  values  of
$uL^{1/\nu}$)  and  confirms  the  hypothesis  in  Eq.~(\ref{Eq:31}).  
Therefore, one can deduce that $\xi/L$ is finite for $\sigma<\sigma_c$
and it decreases when $uL^{1/\nu}$ becomes more negative. In addition,
the  results   confirm  the  compact  character   of  the  subcritical
3D-spanning  avalanche,  as  proposed  in Ref.~\cite{Vives2003}  by  a
different method.

\section{Discussion}
\label{Discussion}

The results  presented so  far together with  the results  obtained in
Ref.~\cite{PerezReche2003}  provide  a clear  scenario  for the  phase
diagram  of  the  $T=0$  3D-GRFIM  with  metastable  dynamics  in  the
thermodynamic limit.

We have  deduced that the subcritical  3D-spanning avalanche occurring
on the transition line given  by Eq.~(\ref{Eq:18.1}) is compact and is
thus  responsible  for  the  macroscopic  jump  of  the  magnetization
\cite{PerezReche2003}. Therefore we  are facing a standard first-order
phase transition scenario with no divergence of the correlation length
for  $\sigma<\sigma_c$.   At  the  critical  point,  this  subcritical
3D-spanning avalanche  becomes fractal at all length  scales, and does
not  fill  any   finite  fraction  of  the  system.    The  end  point
$(\sigma_c,H_c)$ is a standard critical point.

\begin{figure}[ht]
\begin{center}
\epsfig{file=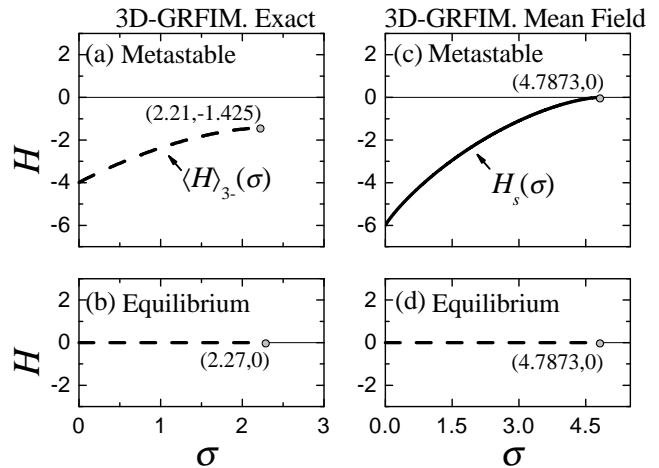, width=8.5cm}
\end{center}
\caption{\label{fig16} Phase diagram corresponding to the exact 3D-GRFIM
  with  metastable dynamics  (a)  and in  equilibrium  (b). The  phase
  diagrams  corresponding  to  the  mean field  approximation  of  the
  3D-GRFIM  with metastable dynamics  (c) and  in equilibrium  (d) are
  also shown. The thick continuous  line in (c) indicates the spinodal
  transition in the  metastable solution and the dashed  lines in (a),
  (b), and (d)  indicate the first-order transitions. In  all cases we
  have indicated the coordinates of the critical point.}
\end{figure}

Fig.~\ref{fig16}(a) shows the obtained phase diagram.  The dashed line
represents    the    first-order     transition    line    given    by
Eq.~(\ref{Eq:18.1}) and  the large dot the critical  point.  Note that
this transition line  is only approximate because it  has been deduced
from scaling  arguments close to the critical  point.  Nevertheless it
is remarkable that Eq.~(\ref{Eq:18.1}) for $\sigma=0$ renders $\langle
H  \rangle_{3-}(0)=-3.999$ which  is unbelievably  close to  the value
$-4$  which can  be computed  by a  (not so)  trivial analysis  of the
coercive field  of the  hysteresis loop of  the Gaussian  3d-RFIM with
metastable    dynamics    corresponding    to   $\sigma    \rightarrow
0^+$.\footnote{A  trivial  analysis for  $\sigma=0$  gives $\langle  H
  \rangle_{3-}(0)=-6$ given that there are 6 nearest-neighbor spins at
  each  site.  Nevertheless,  when  $\sigma \rightarrow  0^+$ one  can
  always  assume that  reversed spins  exist which  act  as nucleation
  sites  for the  subcritical 3D-spanning  avalanche.   This indicates
  that  such an  avalanche will  propagate when  the field  $\langle H
  \rangle_{3-}(0)=-4$ is reached.}

In Fig.~\ref{fig16}(b)  we also show for comparison  the phase diagram
of  the   3D-GRFIM  model  in  equilibrium   at  $T=0$  (ground-state)
\cite{Middleton2002}.  In  addition, in Figs.~\ref{fig16}  (c) and (d)
we  show the  mean  field  (MF) solutions  corresponding  to both  the
metastable  \cite{Sethna1993}   and  equilibrium  \cite{Schneider1977}
cases.

The MF scenario indicates that the equilibrium and metastable critical
points   occur  for   the  same   value   of  $\sigma_c=z\sqrt{2/\pi}$
(Table~\ref{TABLE4}),  where  $z$  is  the  coordination  number.   In
particular, $z=6$  renders $\sigma_c=4.7873$ for  the 3D-GRFIM.  Below
$\sigma_c$, nevertheless, the transition  in equilibrium is a standard
first-order  transition (at  $H=0$),  whereas it  is  a spinodal  line
$H_s(\sigma)$ in  the metastable  case.  From the  equations in  Refs. 
\onlinecite{Sethna1993}  and \onlinecite{Dahmen1996}  it can  be found
that the metastability limit is:
\begin{equation}
\label{Eq:40}
H_s(\sigma)=\sigma
    \sqrt{2\ln \frac{\sigma_c}{\sigma}}
-z\Phi_{err}\left(\sqrt{\ln \frac{\sigma_c}{\sigma}} \right),
\end{equation}
when  the external  field  is decreased.   $\Phi_{err}$  is the  error
function.\footnote{This   error  function   is   normalized  so   that
  $\Phi_{err}(\pm   \infty)=  \pm   1$.}   The   continuous   line  in
Fig.~\ref{fig16}(c)  corresponds  to  $H_s(\sigma)$  for  $z=6$.   The
spinodal  transition   is  characterized   by  a  divergence   of  the
fluctuations  and the  correlation  length on  the line  $H_s(\sigma)$
where the discontinuity of the order parameter $\Delta m$ occurs.

In  the present  work  we have  shown  that when  comparing the  exact
solutions (non  MF) of both  the equilibrium model and  the metastable
model, the character of the  transition line does not change.  In both
cases the transitions are  standard first-order transitions with order
parameter discontinuities  and finite correlation  length. This result
agrees with  the prediction  \cite{Dahmen1996} that the  transition is
abrupt  for $d<8$  as deduced  from an  $\epsilon$  expansion analysis
around $d=8$.

\begin{figure*}[ht]
\begin{center}
\begin{pspicture}(-5,-3)(7,5)
  
\psset{dotsize=6pt 0}

% FOPT surface
 \pspolygon[linecolor=white,fillstyle=gradient,
 gradbegin=darkgray,%
 gradend=white,%
 gradangle=35]%
 (0,0)(3,-1.5)(3.6,-1)(4.2,-0.4)(4.4,-0.15)(4.6,0.2)(4.8,1)(4.9,1.4)(5,2)(0,3)
 \rput{-12}(4,1.7){\large{DS}}

% x-axis (H<0)
  \psline[linewidth=0.75pt](-4,-2)
  \psline[linewidth=0pt,arrowsize=8pt]{->}(-1.2,-0.6)
  \psline[linewidth=0pt,arrowsize=8pt]{->}(-3.2,-1.6)
  \rput[l](-4.6,-2.25){\large{$H$ ($<0$)}}
  
% y-axis (n_max^-1)
  \psline[linewidth=0.75pt](7,0)

% Arrows  
  \rput(6,-0.3){1}
  \rput[l](6.8,0.3){\large{$n_{max}^{-1}$}}

% FOPT. Equilibrium. \sigma<\sigma_c
  \psline[linewidth=2pt,linestyle=dashed](0,3)
  \psline[linewidth=2pt,arrowsize=8pt]{<-}(0,1.8)(0,1.8)
  \psline[linewidth=2pt,arrowsize=8pt]{<-}(0,0.8)(0,0.8)
% FOPT. \sigma>\sigma_c
  \psline[linewidth=0.75pt](0,3)(0,5)
  \psline[linewidth=0pt,arrowsize=8pt]{<-}(0,4.2)(0,3)
  \rput(-0.2,5){\large{$\sigma$}}

% Horizontal line. sigma=0
  \psline[linewidth=1.5pt](3,-1.5)
  \psline[linewidth=1.5pt,arrowsize=7pt]{->}(0.75,-0.375)
  \psline[linewidth=1.5pt,arrowsize=7pt]{->}(2.25,-1.125)

% Position of the Metastable Fixed Point
  \psline[linewidth=0.75pt,linestyle=dotted](5,-0.5)(5,2)

% Line nmax^-1=1, sigma=0
  \psline[linewidth=0.75pt,linestyle=dotted](1,-2.5)(6,0)

% Critical line
  \psline[linewidth=2pt](0,3)(5,2)
  \psline[linewidth=2pt,arrowsize=8pt]{->}(0,3)(1.25,2.75)
  \psline[linewidth=2pt,arrowsize=8pt]{->}(0,3)(3.75,2.25)
  \rput[l]{-11.3}(4.2,2.4){\large{CS}}
  
% Metastable FOPT line
  \pscurve[linewidth=2pt,showpoints=false,linestyle=dashed](5,2)(4.4,-0.15)(3,-1.5)
  \psline[linewidth=0pt,arrowsize=8pt,linestyle=dotted]{->}(4.9,1.4)(4.8,1)
  \psline[linewidth=0pt,arrowsize=8pt,linestyle=dotted]{->}(4.2,-0.4)(3.6,-1)

% FOPT line intermediate
 \pscurve[linewidth=0.75pt,showpoints=false,linestyle=dashed](2.5,2.5)(2.25,0.75)(1.5,-0.75)

% Projection on the H-sigma plane.
  \pscurve[linewidth=1pt,showpoints=false,linestyle=dashed,linecolor=gray](-1,2)(-1.6,-0.15)(-3,-1.5)
  \psset{linecolor=gray}
  \qdisk(-1,2){2pt}
  \psline[linewidth=1pt,linecolor=gray](0,3)(-1,2)
  \psline[linewidth=0.75pt,linestyle=dotted](5,2)(-1,2)
% Projection on the n_max^-1-sigma
  \psline[linewidth=1pt,linecolor=gray](0,3)(6,2.5)
  \qdisk(6,2.5){2pt}
  \psset{linecolor=black}
  \psline[linewidth=0.75pt,linestyle=dotted](6,0)(6,3)
  \psline[linewidth=0.75pt,linestyle=dotted](0,3)(6,3)
  \psline[linewidth=0.75pt,linestyle=dotted](5,2)(6,2.5)

% Dots
  \rput[r](-0.1,3.2){ECP}
  \rput[l](5.2,2){MCP}
  \rput[l](3.2,-1.6){MDFP}
  \rput[r](-0.1,0.2){EDFP}
  \psdots*(0,3)(5,2)
  \psdots*[dotstyle=square*](0,0)(3,-1.5)

% RG Flow on the discontinuity surface
 % \pscurve[linewidth=0.5pt,showpoints=false,arrowsize=6pt]{>->}(4,1)(3.8,0)(3.1,-1.4)
 % \pscurve[linewidth=0.5pt,showpoints=false,arrowsize=6pt]{>->}(1,2.4)(2.4,1.8)(3.2,1.2)(3.4,0)(3.05,-1.4)

  \pscurve[linewidth=0.5pt,showpoints=false,arrowsize=6pt]{>->}(1,2.4)(2.4,1.8)(3.4,1.2)(4,0.6)(4.1,0)(3.05,-1.4)

  \pscurve[linewidth=0.5pt,showpoints=false,arrowsize=6pt]{>->}(1.4,0.2)(2,-0.3)(2.6,-0.8)(3,-1.5)

% RG Flow out of the FOPT surface
  \pscurve[linewidth=0.5pt,showpoints=false,arrowsize=6pt]{>->}(0.8,-0.6)(1.6,-1.2)(1.4,-2)(1,-2.4)

% Legend
%  \rput[l](-4.5,-3.4){CS:   Critical    surface.}
%  \rput[l](-4.5,-4){ECP: Equilibrium   critical  point.}
%  \rput[l](-4.5,-4.6) {MCP:  Metastable critical point.}

%  \rput[l](1,-3.4){DS: Discontinuity surface.}
%  \rput[l](1,-4){EDFP: Equilibrium discontinuity fixed point.}
%  \rput[l](1,-4.6) {MDFP: Metastable discontinuity fixed point.}

%\psgrid 
\end{pspicture}
\end{center}

\caption{\label{fig17} Schematic  phase diagram of  the proposed model
  in the  space $(H,n_{max}^{-1},\sigma)$  in the adiabatic  limit and
  $T=0$.  Arrows  indicate the RG  flow. The meaning of  the different
  acronyms  is explained  in the  text. The  grey lines  correspond to
  projections on the vertical planes.}
\end{figure*}
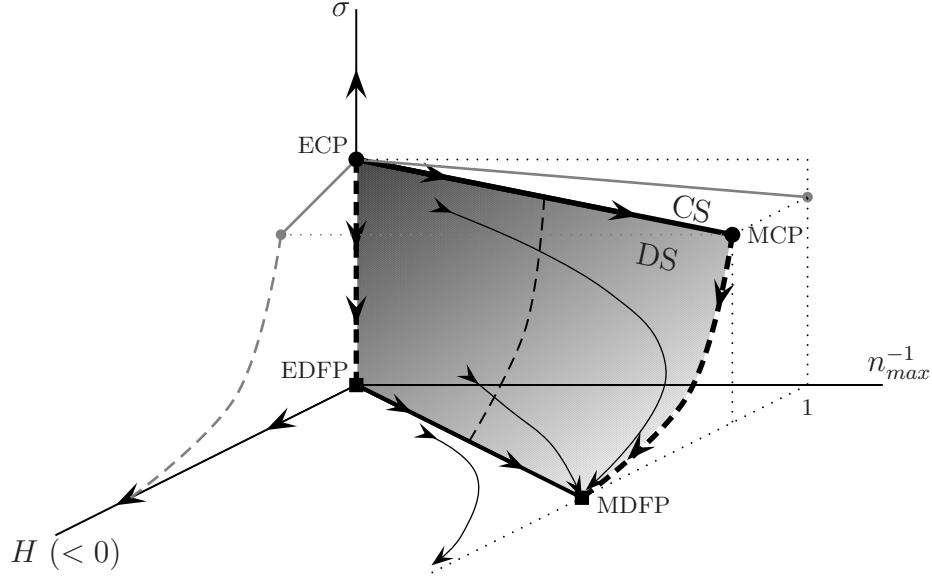

As    indicated    in    Fig.~\ref{fig16}   and    Table~\ref{TABLE4},
$\sigma_{c}^{eq}   =    2.270   \pm   0.004$    in   equilibrium   and
$\sigma_c^{met}= 2.21 \pm 0.02 $ in the metastable case. Regarding the
value  of  the  critical  field,  it is  $|H_c|=1.425$  in  the  exact
metastable model and zero in  the exact equilibrium model.  Thus, when
the exact 3d-GRFIM is studied,  the critical point in equilibrium does
not  coincide   with  that  corresponding  to   metastable  dynamics.  
Nevertheless, the  critical exponents are the  same within statistical
errors.    The   values   are   indicated   in   Table~\ref{TABLE4}.   
\footnote{Our  definition   of  the  exponent   $\mu$  corresponds  to
  $\nu/\beta \delta$ in most of the previous works. We have decided to
  introduce a  new name in order  to emphasize the  importance of this
  exponent which is analogous to  $\nu$ in a RG picture. The exponents
  $\beta$ and $\delta$ are somehow  secondary and take, in the present
  case, different values for the different types of avalanches. In the
  case of  the exact equilibrium  model we have computed  the exponent
  $\mu=\nu/\beta  \delta$  using  the  scaling  relation  $\beta+\beta
  \delta=(d-\tilde{\theta})\nu$    valid    in   equilibrium,    where
  $\tilde{\theta}=1.49 \pm 0.03$  \cite{Middleton2002} is the exponent
  associated with the free energy. We consider that $\beta=\beta_{3-}$
  in our  simulations since the  discontinuity of the  order parameter
  for  $\sigma<\sigma_c$  is related  to  the subcritical  3D-spanning
  avalanche.}  In fact, Dahmen  et al.  \cite{Dahmen1996} have already
pointed out  this similitude between  the critical exponents  for both
models.  These  authors argue that  agreement between the two  sets of
exponents  is  rather  unexpected   since  the  two  models  are  very
different.
\begin{table*}
\begin{center}
\begin{tabular}{|c|cc|cc|}
\hline 
\hline
\multicolumn{1}{|c|}{Magnitude} & \multicolumn{2}{c|}{3D-GRFIM. Mean field} &
\multicolumn{2}{c|}{3D-GRFIM}\\
\cline{2-5}
 & Equilibrium & Metastable & Equilibrium & Metastable\\
 & (Ref.~\cite{Schneider1977})  & (Ref.~\cite{Perkovic1996}) & (Ref.~\cite{Middleton2002})
 & (This work)\\
\hline
$\sigma_c$ & $z\sqrt{2/\pi}$ & $z\sqrt{2/\pi}$ & $2.270 \pm 0.004$ &
$2.21 \pm 0.02$\\
$|H_c|$ & 0 & 0 & 0 & $1.425 \pm 0.010$\\
\hline
$\nu$ & $1/2$ & $1/2$ & $1.37 \pm 0.09$ & $1.2 \pm 0.1$\\
$1/\mu$ & 3 & 3 & $1.498 \pm 0.034$ & $1.5 \pm 0.1$\\
$\beta$ & $1/2$ & $1/2$ & $0.017 \pm 0.005$ & $0.024 \pm 0.012$\\
\hline
\hline
\end{tabular}
\end{center}
\caption{\label{TABLE4} Coordinates of the critical point in the
  $\sigma-H$ plane and critical exponents for the 3D-GRFIM in 
  equilibrium and with metastable dynamics corresponding to the MF
  approximation and the exact models.} 
\end{table*}
Nevertheless,  we can  provide  an argument  based on  renormalization
group theory that indicates that the critical points in the two models
(3D-RFIM  in equilibrium  and  the 3D-RFIM  with metastable  dynamics)
correspond to the same fixed point  in a more general parameter space. 
Within the  framework of RG  theory the critical surface  (or critical
line) is defined as the set  of all points in the parameter space that
flow to a certain critical  fixed point when the renormalization group
transformation is applied. The  variation of the tunable parameters of
a model  describes a ``physical''  trajectory in the parameter  space. 
According to these definitions,  the critical point corresponds to the
point  where  the  ``physical''  trajectory  intersects  the  critical
surface.   The  two  models   discussed  here  can  be  considered  as
particular  cases of  a  more  general model  with  the same  3D-GRFIM
Hamiltonian and  the following $T=0$  adiabatic dynamics: when  $B$ is
varied, blocks of neighboring spins  of size $n \le n_{max}$ flip when
such a  flip represents an  energy decrease.  The  metastable dynamics
introduced  by Sethna  corresponds  to $n_{max}=1$  (only single  spin
flips  are  considered) and  the  equilibrium  model  at $T=0$  (exact
ground-state)   corresponds   to   $n_{max}=\infty$.   The   parameter
$n_{max}$  is  a new  parameter  that must  be  considered  in the  RG
equations.  Since a critical point  is found both with $n_{max}=1$ and
$n_{max}=\infty$, it is  natural to assume that this  is an irrelevant
parameter.    Thus,    we   propose   the    scenario   presented   in
Fig.~\ref{fig17}.   Changing  $n_{max}$  alters  the position  of  the
critical point, but not the critical exponents which correspond to the
same critical fixed point.  Numerical simulations of the 3D-GRFIM with
$n_{max}>1$ dynamics will help  in clarifying this picture. At present
we guess that  the RG flow follows the  arrows schematically indicated
in Fig.~\ref{fig17}.   Both the  equilibrium critical point  (ECP) and
the metastable critical  point (MCP) lie on the  same critical surface
(CS). In general  a first-order phase transition occurs  at the points
in the  parameter space  that go towards  a discontinuity  fixed point
when the  RG transformation is  applied.~\cite{Nienhuis1975} We assume
the  existence  of two  discontinuity  fixed  points: the  equilibrium
discontinuity  fixed  point (EDFP)  and  the metastable  discontinuity
fixed  point   (MDFP).   The  EDFP  controls   the  first-order  phase
transition  in the  equilibrium case  ($n_{max}^{-1}=0$ and  $\sigma <
\sigma_{c}^{eq}$)  and   the  MDFP  controls   the  first-order  phase
transition when  $n_{max}^{-1}>0$.  All  the points that  flow towards
any of the discontinuity fixed points define the discontinuity surface
(DS) where the first-order phase transition occurs.

Another interesting  question to be discussed is  the determination of
the correlation length $\xi$ in the 3D-GRFIM with metastable dynamics.
Avalanches can  be understood as the zero  temperature fluctuations in
the driven system. Is their average linear size related to $\xi$?. The
first thing to note is that  we have found that avalanches display two
different  fractal dimensions (and  thus different  associated $\beta$
exponents). The $\nu$ and  $\mu$ exponents, nevertheless, are the same
for all the  scaling collapses.  For instance, this  is illustrated by
Fig.~\ref{fig7}  in  this  work  and   by  Figs.   8,  9,  and  10  in
Ref.~\cite{PerezReche2003}.   Thus  the  behavior of  the  correlation
length is unique:
\begin{equation}
\xi = u^{-\nu} \hat{\Xi}(u^{\nu}v^{-\mu}),
\end{equation}
with $\hat{\Xi}(x)  \sim x+ {\cal O}(x^2)$. The  fluctuations then can
``choose''  between two  different mechanisms  for  propagation either
with  fractal   dimension  $d_{f}=2.78$  or   with  fractal  dimension
$d_{3-}=2.98$. The second point to  be considered is that $\xi$ cannot
be related to the size  of the subcritical 3D-spanning avalanche since
we have found that $\xi$ is finite below $\sigma_c$. Keeping these two
observations in mind we propose that the correlation length is related
to the  average radius of  the largest non-spanning  avalanche.  Below
$\sigma=\sigma_c$ the  existence of a  compact subcritical 3D-spanning
avalanche does not allow for the non-spanning avalanches to overcome a
certain finite length and thus  $\xi$ is finite.  Only at the critical
point does  the subcritical  3D-spanning avalanche become  fractal and
allows  for  other spanning  avalanches  to  exist  and $\xi$  becomes
infinite.  This  behavior is  much similar to  what has  been recently
found in  percolation.  \cite{Aizenman1997,Stauffer1997} We conjecture
that some of the theorems  that have been rigorously proven concerning
the  uniqueness   of  the  infinite  percolating   cluster  should  be
applicable to our case  concerning the compact subcritical 3D-spanning
avalanche. The present results  should be considered as an interesting
stimulus  to proceed  with the  analysis of  percolation  theory.  For
instance,  we propose checking  whether the  fractal dimension  of the
spanning  clusters is  the same  as that  of the  infinite  cluster at
distances  lower  than  the  correlation  length  at  the  percolation
threshold.

\section{Summary and conclusions}
\label{Conclusions}
The results presented  in this paper are mainly  related to two topics
in  the  3D-GRFIM:  firstly,  the  field dependence  of  the  spanning
avalanches,   and  secondly,   the  geometrical   properties   of  the
avalanches.   We   have  extended  the  FSS   hypothesis  proposed  in
\cite{PerezReche2003}  to   properly  take  into   account  the  field
dependence of the number densities $n_{\alpha}(H,\sigma,L)$ and of the
bivariate   distributions  ${\cal   D}_{\alpha}(s,H;\sigma,L)$.   When
carrying out  such an  extension, it is  necessary to introduce  a new
scaling  variable  $v$  and  a  new  exponent  $\mu$  related  to  the
divergence of  the correlation length when $H$  approaches $H_c$ ($\xi
\sim (H-H_c)^{-\mu}$).   We have also introduced  a scaling hypothesis
for  the field  $\langle  H \rangle_{\alpha}(\sigma,L)$  at which  the
different   avalanches  concentrate   and  their   standard  deviation
$\sigma_{\alpha}^H(\sigma,L)$.     From    the    scaling    collapses
corresponding  to the  1D- and  2D-spanning avalanches  we  have found
$1/\mu  =1.5$.   The  study  of  the 3D-spanning  avalanches  is  more
intricate as already shown  in \cite{PerezReche2003}, where we propose
the  existence of two  different kinds  of 3D-spanning  avalanches. In
this  paper we have  proposed two  approximate separation  methods for
classifying these avalanches as  subcritical or critical.  Using these
methods we have found that $1/\mu=1.5$ for both cases. Scaling enables
the following behavior to be scketched in the thermodynamic limit: The
1D-,  2D-,  and critical  3D-spanning  avalanches  only  exist at  the
critical  point  $(\sigma_c,H_c)=(2.21,-1.425)$,  where  their  number
densities  are  infinite.  In  contrast,  one subcritical  3D-spanning
avalanche exists below $\sigma_{c}$ and it occurs on the line $\langle
H \rangle_{3-}(\sigma)$ (Eq.~(\ref{Eq:18.1})).

From the  average mass $M_{\alpha}(\ell;\sigma)$ we  have obtained the
fractal  dimensions corresponding  to each  of the  types  of spanning
avalanches. This  has allowed us to confirm  independently the results
in \cite{PerezReche2003}:  $d_f=2.78 \pm 0.05$  for the 1D-,  2D-, and
subcritical  3D-spanning  avalanches   and  $d_{3-}=2.98  \pm  0.02$.  
Furthermore,   the  behavior   for  $\sigma<\sigma_c$   of   the  mass
corresponding to the subcritical 3D-spanning avalanches indicates that
the  correlation  length  $\xi$  is  finite below  $\sigma_c$.   As  a
consequence,    we    conclude    that    the    line    $\langle    H
\rangle_{3-}(\sigma)$, where the  discontinuity in the order parameter
occurs, corresponds  to a  standard first-order phase  transition line
and $\xi$ only diverges at the critical point.

\section*{Acknowledgements}
We   acknowledge  fruitful   discussions  with   X.Illa,  B.Tadi\'{c},
Ll.Ma\~nosa, and  A.Planes.  This work has  received financial support
from  CICyT  (Spain), project  MAT2001-3251  and  CIRIT (Catalonia)  ,
project  2000SGR00025. This  research  has been  partially done  using
CESCA  resources   (project  CSICCF).   F.J.    P.  also  acknowledges
financial support from DGICyT.

\end{document}